\documentclass[journal]{IEEEtran}
\pdfoutput=1
\usepackage{enumerate}
\usepackage{makeidx}  
\usepackage{graphicx}
\usepackage{amsmath}
\usepackage{subfigure}
\usepackage{slashbox}
\usepackage{color}
\usepackage[english]{babel}

\begin{document}
%
\title{Geo-distinctive Visual Element Matching for Location Estimation of Images}

\author{Xinchao~Li,~
        Martha~Larson,~
        Alan~Hanjalic~\IEEEmembership{Fellow,~IEEE}
\thanks{X. Li, M. Larson and A. Hanjalic are with the Multimedia Computing Group, Delft University of Technology, Delft, The Netherlands.
 e-mail: \{x.li-3,m.a.larson,a.hanjalic\}@tudelft.nl.}
}

\maketitle

\begin{abstract}
We propose an image representation and matching approach that substantially improves visual-based location estimation for images. 
The main novelty of the approach, called distinctive visual element matching (DVEM), is its use of representations that are specific to the query image whose location is being predicted.
These representations are based on visual element clouds, which robustly capture the connection between the query and visual evidence from candidate locations. 
We then maximize the influence of visual elements that are geo-distinctive because they do not occur in images taken at many other locations.
We carry out experiments and analysis for both geo-constrained and geo-unconstrained location estimation cases using two large-scale, publicly-available datasets: the San Francisco Landmark dataset with $1.06$ million street-view images and the MediaEval '15 Placing Task dataset with $5.6$ million geo-tagged images from Flickr.  
We present examples that illustrate the highly-transparent mechanics of the approach, which are based on common sense observations about the visual patterns in image collections.
Our results show that the proposed method delivers a considerable performance improvement compared to the state of the art.
\end{abstract}

%
\IEEEpeerreviewmaketitle

\section{Introduction}
\label{sec:intro}
Information about the location at which an image was taken is valuable image metadata.
Enriching images with geo-coordinates benefits users by supporting them in searching, browsing, organizing and sharing their images and image collections. 
Specifically, geo-information can assist in generating visual summaries of a location~\cite{visSum_Stevan2013,LocVisDiversity2013}, in recommending travel tours and venues~\cite{TravelRecommendation, VenueRecommendation15}, in photo stream alignment~\cite{phoStreamAlign}, and in event mining from media collections~\cite{EventMining_2010,Choi2015Event}.

While many modern mobile devices can automatically assign geo-coordinates to images during capture, a great number of images lack this information~\cite{Serdyukov:2009:PlacingFlickrPhotos}. 
Techniques that automatically estimate the location of an image~\cite{Crandall:2009:MappingWorldPhotos,Serdyukov:2009:PlacingFlickrPhotos,IM2GPS,MobileVisLoc13,Martha:2011:TaggingGeoTagging} have been receiving increasing research attention in recent years.
Specifically, predicting geographic location solely from visual content holds the advantage of not depending on the availability of the textual annotation. 
The challenge of visual content-based geo-location estimation derives from the relationship between visual variability and location.
Images taken at a single location may display high visual variability, whereas images taken in distinct locations may be unexpectedly similar.

\begin{figure}[t]
  \centering
  \scalebox{0.48}{\includegraphics[width=\textwidth]{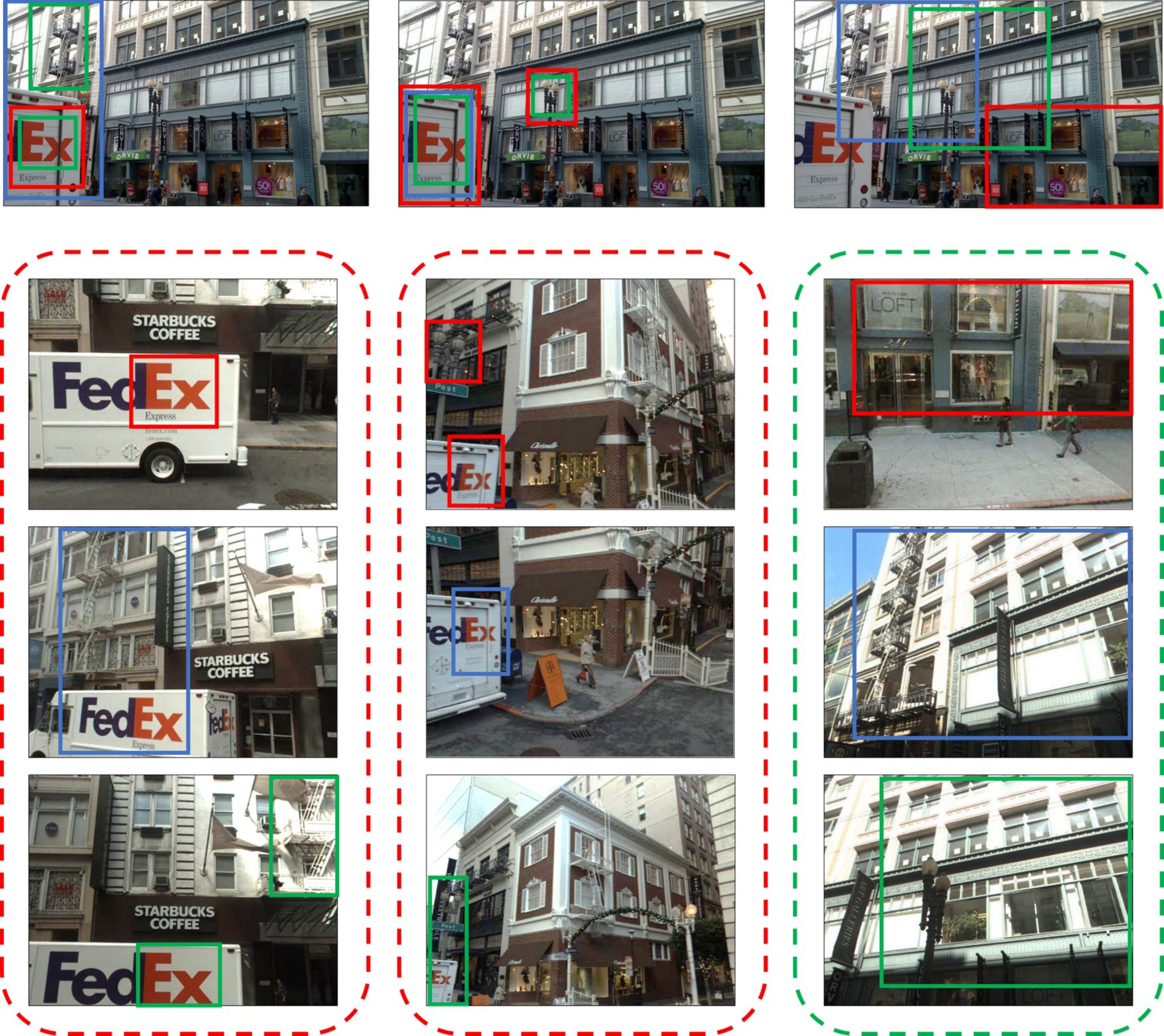}}
  \vspace{-0.5em}
  \caption{Colored boxes indicate match between areas of a query image (top row) and location images taken at three different locations (columns). Note how these areas differ when the location is different from the query location (left and middle columns) and when it is the same (right column).}\label{fig:exam_coreIdea}
\end{figure}

The core idea underlying our approach to this challenge is depicted in Fig.~\ref{fig:exam_coreIdea}, which illustrates the pattern of visual matching that we will exploit in this paper.
Inspecting each column of images in turn, we can see similarities and differences among the areas of the images marked with colored boxes.
These areas contain visual elements that match between query image (top row) and the location images (lower rows).
We use the term \emph{visual element} to denote a group of pixels (i.e., an image neighborhood) that is found around salient points and that also can automatically be identified as being present in multiple images, i.e., by means of visual matching. 

Moving from left to right in the figure, we notice that the areas matched in the first two locations (left and middle columns) share similarity. 
Here, the visual elements contained in these areas correspond to FedEx trucks, street lights, and fire escapes.
The locations in these two columns are \emph{different} from the query location. 
These visual matches introduce visual confusion between the query image and images taken at other locations.
In contrast, location in the third column is the \emph{same} as the query location.
The matching areas contain visual elements correspond to specific, distinguishing features of the real-world location, not found in other locations, in this case, elements of the architecture.
We call such visual elements \emph{geo-distinctive}.

This paper introduces a visual matching approach to image geo-location estimation that exploits geo-distinctive visual elements, referred to as \emph{distinctive visual element matching} (DVEM).
This approach represents a contribution to the line of research dedicated to developing search-based approaches to visual-content-based geo-location estimation for images.
Under search-based geo-location estimation, the target image (whose geo-coordinates are unknown) is used to query a \emph{background collection}, a large collection of images whose geo-coordinates are known.
Top-ranking results from the background collection are processed to produce a prediction of a location, which is then propagated to the target image.
As is customary in search-based approaches, we refer to the target image as the \emph{query image}.
The DVEM approach represents a significant extension to our generic \emph{geo-visual ranking} framework~\cite{my_TMM_GVR} for image location estimation.

As will be explained in detail in Section~\ref{Rationale} and~\ref{relatedWork}, DVEM represents a considerable advancement of the state of the art in search-based approaches to visual-content-based image geo-location estimation.
In a nutshell, the innovation of DVEM is its use of a visual representation that is `contextual' in that it is specific to the query image.
This representation is computed in the final stage of search-based geo-location, during which top-ranked results are processed.
The key is that the representation is not fixed in advance, but rather is calculated at prediction time, allowing it to change as necessary for different queries.
This factor sets DVEM apart from other attempts in the literature to exploit geo-distinctiveness, which pre-calculate representations based on the background collection, rather than zeroing in on visual information most important for an individual query.
The experimental results we present in this paper demonstrate that DVEM can achieve a substantial improvement for both major types of image geo-location prediction covered in the literature: geo-constrained and geo-unconstrained. 

The remainder of the paper is organized as follows. 
In Section II, we present the rationale underlying our proposed approach, DVEM, and describe its novel contribution in more detail. 
Then, in Section~\ref{relatedWork}, we provide an overview of the related work in the domain of image location estimation and position our contribution with respect to it. 
Section~\ref{Sec_DVEM} describes the DVEM approach in detail. 
Our experimental setup is explained in Section~\ref{ExperFrame} and Section~\ref{ExperRes} reports our experimental results.
Section~\ref{conclusion} concludes the paper and provides an outlook towards future work.

\section{Rationale and Contribution}\label{Rationale}
The fundamental assumption of content-based geo-location estimation is that two images that depict the same objects and scene elements, are likely to have been taken at the same location.
On the basis of this assumption, search-based geo-location estimation exploits image content by applying object-based image retrieval techniques. 
The rationale for our approach is grounded in a detailed analysis of the particular challenges that arise when these techniques are applied to predict image location. 

We examine these challenges in greater depth by returning to consider Fig.~\ref{fig:exam_coreIdea}. 
In Section~\ref{sec:intro}, we have already discussed the existence of confounding visual elements in images from the wrong location (left and middle columns), and also of characteristic visual elements in images from the true location (right column). 
We now look again at these cases in turn.

\noindent\textbf{Geo-distinctivness} Images taken at a wrong location (Fig.~\ref{fig:exam_coreIdea} left and middle) capture a underlying reality that is different from the reality captured by the query.
The figure shows two typical sources of confounding visual elements.
First, elements corresponding to real-world objects that are able to move from one location to the other, such as a FedEx truck.
Second, elements corresponding to objects that are identical or highly similar and occur at multiple locations, such as the fire escapes and the street lamps.
A third case (not depicted) occurs when objects or scene elements at different locations appear having similar visual elements in images due to the way in which they were captured (i.e., perspective, lighting conditions, or filters).

Our approach is based on the insight that confounding visual elements will occur in many locations that are \emph{not} the true location of the image. 
DVEM is designed to limit the contribution of visual elements that occur in many locations, and instead base its prediction on visual elements that are discriminative for a specific location.

\noindent\textbf{Location representation} Images taken at the true location (Fig.~\ref{fig:exam_coreIdea} right column) imply a related set of challenges.
Conceptually, to relate a query image and its true location, we would like to count how many visual elements in the query correspond to real-world aspects of the location.
Practically, however, such an approach is too na\"{i}ve, since we cannot count on our image collection to cover each location comprehensively.
Further, we face the difficulty that the true-location images in our background collection may have only a weak link with the query image.
Specifically for the example in Fig.~\ref{fig:exam_coreIdea}, the variation in the perspective is significant between the query and the images from the true location (right column), which will heavily weaken their visual correspondences. 
We again must deal with the same set of factors that give rise to confounding visual elements, mentioned above: camera angle, zoom-level, illumination, resolution, and filters.
These also include the presence of mobile objects such as pedestrians, vehicles, and temporary signs or decorations.
We have no control over the presence of these distractors, but we can seek to reduce their impact, which will in turn limit their contribution to the match between query and wrong locations.

DVEM builds on the practical insight that we should focus on aggregating evidence strength, rather than merely counting visual elements common between a query and a location.
In particular, we aim to integrate two tendencies, which are illustrated by the right column of Fig.~\ref{fig:exam_coreIdea}.
Here, it can be see that the match between query image and true location involves (a) a wider variety of different visual elements than matches with wrong locations and (b) visual elements that are distributed over a larger area within the image.
These tendencies can be considered to be reflections of the common sense expectation that the number of ways in which a query can overlap with true-location images is much larger than the number of ways in which a query can overlap with wrong-location images.

\noindent\textbf{Connection with search-based geo-location estimation} Next we turn to describe how DVEM extends our general \emph{geo-visual ranking} (GVR) framework~\cite{my_TMM_GVR}. 
\begin{figure*}[t]
  \centering
  \scalebox{0.9}{\includegraphics[width=\textwidth]{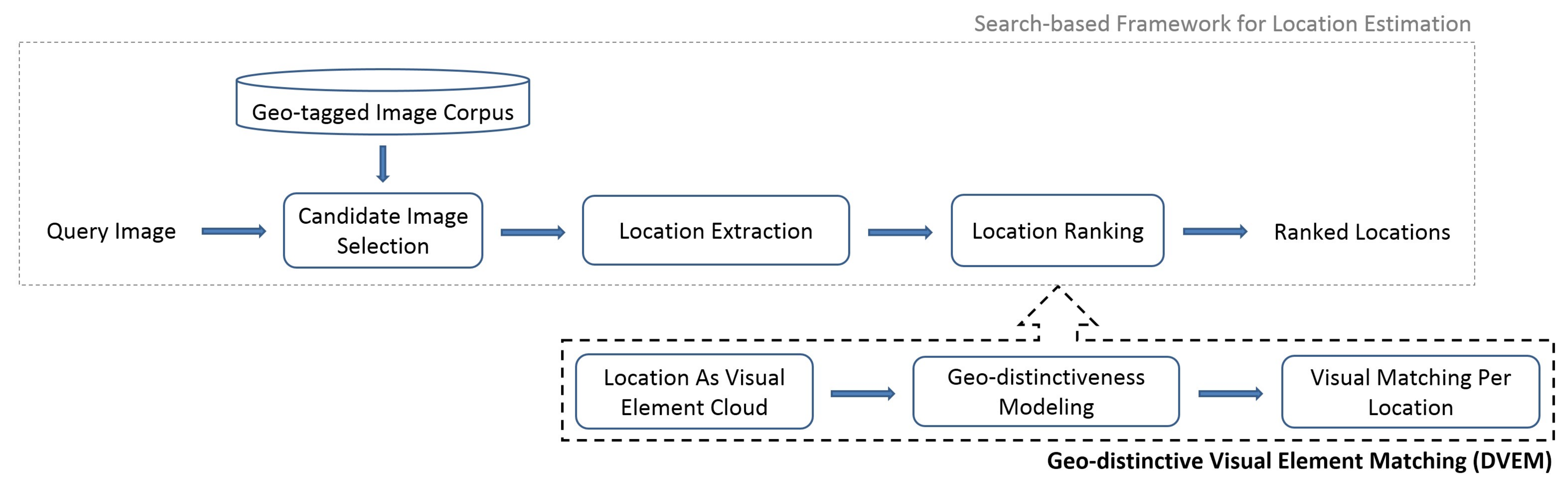}}
  \caption{Our propsed \emph{Geo-distinctive Visual Element Matching} (DVEM) approach, depicted with its integration as the location ranking step of the generic search-based location estimation framework~\cite{my_TMM_GVR}.  }\label{fig:SysScheme}
\end{figure*}
As previously mentioned, DVEM contributes to the processing step in a search-based geo-location estimation pipeline. 
Fig.~\ref{fig:SysScheme} depicts the GVR framework in the top row, and the DVEM extension in the bottom row. 
The dashed line indicates the steps that compose DVEM and the arrow show that it replaces the Location Ranking step of GVR.

Here, we provide a brief review of the functioning of GVR.
In the Candidate Image Selection step, we use the query image to query a background collection (corpus) of geo-tagged images, i.e., images annotated with geo-coordinates.
In the Location Extraction step, we group the retrieved images according to their locations, creating image sets corresponding to candidate locations.
This information serves as input into DVEM.

The three steps of DVEM are designed to address the challenges covered at the beginning of the section, and incorporate both geo-distinctiveness and location representation:
\begin{itemize}
\item \emph{Location as Visual Element Cloud} builds a `contextual' query-specific representation of each candidate-location image set that reflects the strength of the visual evidence relating that image set to the query.
\item \emph{Geo-Distinctiveness Modeling} captures the ability of visual elements to discriminate the image sets of the candidate locations that are competing for a given query.
\item \emph{Visual Matching per Location} calculates the ranking score for each candidate location with the target to incorporate both the distinctiveness of visual elements and the matching strength between visual elements and the location.
\end{itemize}
These steps are explained in detail in Section~\ref{Sec_DVEM}, which also includes further motivating examples.

\noindent\textbf{Novel contributions} As stated in the introduction, the novel contribution of DVEM is its use of query-specific, `contextual', visual representations for geo-location estimation. 
No collection-wide representation of location is needed.
Instead, flexible representations are built at prediction time that aggregate evidence for ranking a location optimally against its specific competitors for each query. 
The implications of this contribution are best understood via a comparison with classical information retrieval.
DVEM can clearly claim the traditional vector space model with TF-IDF weighting scheme used in information retrieval as a progenitor. 
TF-IDF consists of a Term Frequency (TF) component, which represents the contents of items (documents), and an Inverse Document Frequency (IDF) component, which discriminates items from others in the collection~\cite{IRBook1999}. 
DVEM uses the same basic principle of combining a representative component, the visual element cloud, and a discriminative component, geo-distinctiveness modeling. 
However, its application of these principles is unique, and differentiate DVEM from the ways in which TF-IDF has been deployed for bag-of-feature-based image retrieval in the past.

\begin{itemize}
\item DVEM moves matching from the level of the item (i.e., individual image), to the level of the candidate image set. The visual element cloud generated from the candidate image set makes it possible for individual visual elements to contribute directly to the decision, compensating for the potentially weak visual link of any given location image with the query.
\item DVEM dispenses with the need to define individual locations at the collection level offline at indexing time. Instead DVEM defines `contextual' visual representations of locations over the candidate image sets, which represent the images most relevant for the decision on the location of a particular query at prediction time.
\end{itemize}

The use of `contextual' visual representations of locations that are created specifically for individual queries have two important advantages.
First, these representations involve only images that have been visually verified in the candidate image selection step.
Since images that are not relevant to the location estimation decision are not present in the candidate image set, the location representations can focus on the `contextual' task of ranking the competing locations to make the best possible decision for a given query, improving robustness.
Second, the number of competing locations for any given query is relatively low, meaning that the geo-distinctiveness calculation is computationally quite light. This solves the problem of making geo-distinctiveness computationally tractable. It allows DVEM to scale effortlessly as the number of possible candidate locations grows to be effectively infinite in the case of geo-location estimation at global scale.

As we will show by experimental results in Section~\ref{ExperRes}, these advantages delivers an overall significant improvement of the location estimation performance compared to state-of-the-art methods. 

\section{Related Work}\label{relatedWork}
Visual-only geo-location estimation approaches can be divided into two categories.
The first is \emph{geo-constrained} approaches.
Such approaches estimate geo-location within a geographically constrained area~\cite{landmark_identification2011,repetStru_CVPR13} or a finite set of locations~\cite{PlacesOfInterest_TMM13,imagRetri_localization:2010,large-scaleLandMark:2009,Landmark3D}. 
The second is \emph{geo-unconstrained} approaches, which estimate geo-location at a global scale~\cite{IM2GPS,my_TMM_GVR}. 
The challenge of geo-unconstrained geo-location estimation is daunting: a recent survey~\cite{larson14} indicated that there are still ample opportunities waiting to be explored in this respect.
In this work, our overall goal is to substantially improve the accuracy of image location estimation using only their visual content, and to achieve this improvement in both the geo-constrained and geo-unconstrained scenarios. 
As demonstrated by our experimental results, DVEM's representation and matching of images using geo-distinctive visual elements achieves a substantial performance improvement compared to existing approaches to both geo-constrained and geo-unconstrained location estimation.

\subsection{Geo-constrained content-based location estimation}\label{relatedWork_geoConstrain}
\noindent\textbf{City-scale location estimation.} Chen et al.~\cite{landmark_identification2011} investigated the city-scale location recognition problem for cell-phone images. 
They employed a street view surveying vehicle to collect panoramic images of downtown San Francisco, which were further converted into 1.7 million perspective images. 
Given a query image taken randomly from a pedestrian's perspective within the city, a vocabulary-tree-based retrieval scheme based on SIFT features~\cite{SIFT} was employed to predict the image's location by propagating the location information from the top-returned image. 

Gopalan~\cite{SpareseCodingGeoLoc_CVPR15}, using the same data set, modeled the transformation between the image appearance space and the location grouping space and incorporated it with a hierarchical sparse coding approach to learn the features that are useful in discriminating images across locations. 
We choose this dataset for our experiments on the geo-constrained setting, and use this approach as one of our baselines. The other papers that evaluate using this data set are the aggregated selective matching kernel purposed by Tolias et al. (2015)~\cite{ASMK_IJCV15}, 
the work exploiting descriptor distinctiveness by Arandjelovi\'{c} and Zisserman (2014)~\cite{DisLocation2014}, the work exploiting repeated pattens by Torii et al. (2013)~\cite{repetStru_CVPR13}, 
the graph based query expansion method of Zhang et al. (2012)~\cite{zhang2012Qfusion} and the initial work of Chen et al. (2011)~\cite{landmark_identification2011}. 
The experiments in Section~\ref{compStateOfArt} makes a comparison with all of these approaches. 

The DVEM is suited for cases in which there is no finite set of locations
to apply a classification approach.
However, we point out here, that classification approaches have been proposed for geo-constrained content-based location estimation.
Gronat et al.~\cite{PerLocationSVM} modeled each geo-tagged image in the collection as a class, and learned a per-example linear SVM classifier for each of these classes with a calibration procedure that makes the classification scores comparable to each other. 
Due to high computational cost in both off-line learning and online querying phases, the experiment was conducted on a limited dataset of $25k$ photos from Google Streetview taken in Pittsburgh, U.S., covering roughly an area of $1.2 \times 1.2 km^2$. 

\noindent\textbf{Beyond city scale.} 
Authors that go beyond city scale, may still address only a constrained number of locations.
Kalantidis et al.~\cite{imagRetri_localization:2010} investigate location prediction for popular locations in $22$ European cities using \emph{scene maps} built by visually clustering and aligning images depicting the same view of a scene.
Li et al.~\cite{PlacesOfInterest_TMM13} constructed a hierarchical structure mined from a set of images depicting about 1,500 predefined places of interest, and proposed a hierarchical method to estimate image's location by matching its visual content against this hierarchical structure.
Our approach resembles~\cite{imagRetri_localization:2010} in that we also use sets of images to represent locations. 
Note however that in DVEM location representations are created specifically for individual queries at prediction time, making it possible to scale beyond the fixed set of locations. 

\subsection{Geo-unconstrained content-based location estimation}\label{relatedWork_geoUnConstrain}
Estimating location from image content on a global scale faces serious challenges.
First, there is effectively an infinite number of locations in the world. 
Second, geo-unconstrained location prediction is generally carried out on large collections of user-contributed social images. 
As a consequence, less photographed locations are underrepresented. 
These challenges imply that geo-unconstrained location estimation cannot be addressed by training a separate model for each location.
Finally, the visual variability of images taken a given location is often high, and is also quite erratic.
For instance, images taken at a location of a monument that is a tourist attraction will probably focus on some aspects of the monument, limiting the scope of the captured visual scene. 
However, images taken at an arbitrary beach may be taken from any view point to capture a wide variety of the visual scene. 
This variability can heavily hinder inference of location-specific information from the visual content of images, and exacerbates the difficulty of linking images showing different aspects of a location. 

The problem of geo-unconstrained content-based image location estimation was first tackled by Hays and Efros~\cite{IM2GPS}. 
They proposed to use visual scene similarity between images to support location estimation with the assumption that images with higher visual scene similarity were more likely to have been taken at the same location. 
In recent years, research on geo-unconstrained location prediction has been driving forward by the MediaEval Placing Task~\cite{larson14}.
The Placing Task result most relevant to DVEM is our submission to the 2013 Placing Task~\cite{my_MedEvl13}.
This submission deployed a combination of local and global visual representations within the GVR system~\cite{my_GVR_ICMR13}, and out-performed other visual-content-based approaches that year.
Here, we will focus on 2015, the most recent edition of the Placing Task~\cite{PlacingWorkNote2015}, which received three submissions using visual-content-based approaches.
Kelm et al.~\cite{IMCUBE_MedEvl15} exploited densely sampled local features (pairwise averaged DCT coefficients) for location estimation. 
Since this submission is not yet a functional, mature result, it is not considered further here. 
Li et al.~\cite{RECOD_MedEvl15} employed a rank aggregation approach to combine various global visual representations in a search-based scheme, and used the top ranked image as the source for location estimation. 
Instead of using hand-crafted features, Kordopatis-Zilos et al.~\cite{CERTH_MedEvl15} made use of the recent developments in learning visual representations. 
They fed a convolutional neural network with images from $1,000$ points of interest around the globe and employed it to generate the CNN features. Location is then estimated for the query image by finding the most probable location among the most visually similar photos calculated based on their proximity in the feature space. 

Our DVEM is related to these approaches in the sense that it is data driven and search based. 
However, these approaches are dependent on finding significant image-level matches between the query image and individual images in the background collection. 
They do not attempt to compensate for the possibility that the match between the query image and individual images taken at the true location might be minimal, due to the way in which the image was taken, or exploit geo-distinctiveness.

\subsection{Geo-distinctive visual element modeling}\label{relatedWork_disticntive}

As discussed in Section~\ref{Rationale}, in a classical information retrieval system, document (item) distinctiveness is traditionally computed off-line during the indexing phase at the level of the entire collection~\cite{IRBook1999}. 
This approach is also used in the classical bag-of-feature-based image retrieval system. 
For example, in~\cite{videoGoogle}, the distinctiveness of each visual word is generated from its distribution in the image database. 
Note that our system uses the approach of~\cite{videoGoogle} in the Candidate Image Selection step (first block in Fig.~\ref{fig:SysScheme}), as a standard best practice. 
Our novel use of geo-distinctiveness goes above and beyond this step, as described in the following.

The key example of the use of distinctiveness for content-based geo-location estimation is the work of Arandjelovi\'{c} and Zisserman~\cite{DisLocation2014}, who modeled the distinctiveness of each local descriptor from its estimated surrounding local density in the descriptor space. This approach differs from ours in two ways: first, we use geo-distinctiveness, calculated on the basis of individual locations, rather than general distinctiveness and, second, we use geo-metrically verified salient points, rather than relying on the visual appearance of the descriptors of the salient points.
As we will show by experimental results in Section~\ref{ExperRes}, which uses Arandjelovi\'{c} and Zisserman~\cite{DisLocation2014} as one of the baselines, this added step of geo-distinctive visual elements matching significantly improves location estimation

Where geo-distinctivenss has been used in the literature, it has been pre-computed with respect to image sets representing a pre-defined inventory of locations.
Typical for such approaches is the work from Doersch et al.~\cite{WhatMakesParis2012Graphics}. 
They built a collection of image patches from street view photos of 12 cities around the world, and mined the image patches that are location-typical---both frequent and discriminative for each city---based on the appearance similarity distribution of the image patches. 
Similarly, Fang et al.~\cite{fang2013giant} incorporated the learned geo-representative visual attributes into the location recognition model in order to improve the classification performance. 
These learned geo-representative visual attributes were shown to be useful for city-based location recognition, i.e., to assign a given image to one of the cities. 
However, this approach faces a significant challenge. 
As the number of pre-defined locations grows larger, there less geo-representative
visual attributes exist per location.
For this reason, the risk increases that an query image contains few of the location-typical elements that have been found for the location at which it was taken.
In our work, instead of extracting location-typical features from the image collection and using them to assess the query, we turn the approach around.
Our approach is to focus on the visual elements that we extract from the query, and to model their geo-distinctiveness around candidate locations for this particular query at prediction time. 

\section{Geo-distinctive visual element matching}\label{Sec_DVEM}
In this section, we present DVEM in depth, providing a detailed description of the components depicted in Fig.~\ref{fig:SysScheme}.
We start with the GVR framework~\cite{my_TMM_GVR} (Fig.~\ref{fig:SysScheme}, top row), the generic search-based location estimation pipeline upon when DVEM builds.
The framework was described in Section~\ref{Rationale}.
 Here, we provide the necessary additional detail.

The first step of GVR is Candidate Image Selection, and serves to retrieve, from the collection of geo-tagged images, a ranked list of candidate images that are most visually similar to the query $q$. 
In contrast to the original version of GVR, our new pair-wise geometrical matching approach is used for this step~\cite{my_CVPR15_PGM}.
The result is a ranked list of images that have been visually verified, ensuring that we can be confident that their visual content is relevant for the decision on the location of the query image.
We limit the ranked list to the top $1000$ images, since this cutoff was demonstrated to be effective in~\cite{my_TMM_GVR}. 
In the second step, Location Extraction, candidate locations are created by applying a geo-clustering process to the ranked list (see~\cite{my_TMM_GVR} for details), resulting in the set $G$ of candidate locations. 
The set of images $I_g$ associated with each location $g$ in $G$ is referred to as the \emph{candidate location image set}.
In the third step, Location Ranking, visual proximities for each $g$ are calculated on the basis of sets $I_g$ and the query $q$, resulting in $Score(g,q)$.
Finally,  $Score(g,q)$ is used to rank the locations $g$ in $G$.
The top-ranked location provides the geo-location estimate, and is propagated to the query image. 
As previously mentioned, DVEM replaces the Location Ranking step of GVR. Specifically, it contributes an advanced and highly effective method for calculating $Score(g,q)$.
The remainder of this section discusses each of the steps of DVEM (bottom row Fig.~\ref{fig:SysScheme}) in turn.

\begin{figure*}[t]
  \centering
  \scalebox{0.95}{\includegraphics[width=\textwidth]{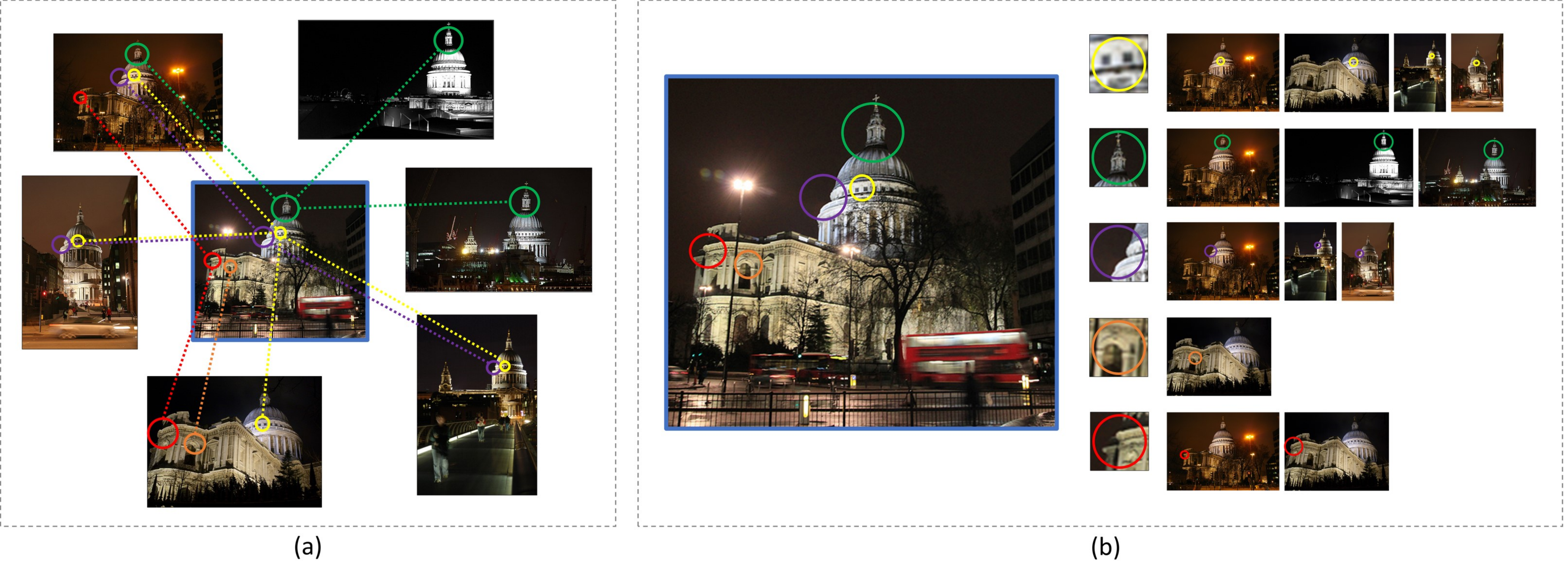}}
  \caption{Illustration of the visual element cloud. Figure (a) shows the correspondences $c$ between the query image (center) and images taken in one location. The relationship between the visual element cloud constructed for this location and the query is illustrated in Figure (b). The cloud is represented by the visual elements from the query and the images of the location these elements appear in}\label{fig:exam_LocVisClould}
\end{figure*}

\subsection{Location as visual element cloud}\label{SubSec_LocVisCloud}
The visual element cloud is a representation of $I_g$ that aggregates the evidence on the strength of the visual link between $I_g$ and the query $q$.
The cloud, illustrated in Fig.~\ref{fig:exam_LocVisClould}, serves as a representation of the location $g$ in terms of visual elements that occur in the query.
For the first step of creating the cloud, we adopt the standard approach (as used, e.g., with SIFT) of detecting salient points in the images using a salient point detector and representing these points with feature vectors (i.e., descriptors) describing the local image neighborhoods around the points. 
The size of the neighbor is determined by the salient point detector.

Next, we calculate correspondences between the salient points in the query and in the individual images on the basis of the local image neighborhoods of the points.
Then, we apply geometric matching, which secures the consistency of transformation between different salient points. 
In this work, we use PGM~\cite{my_CVPR15_PGM}, as applied in the Candidate Image Selection step, but another geometric verification approach could also be chosen.
The result of geometric matching is a set of one-to-one correspondences $c$ between salient points in the query and in the individual images $I_g$ (cf. Fig.~\ref{fig:exam_LocVisClould}a), and a set of matching scores $IniScore(c)$ associated with the correspondences $c$.
The \emph{visual elements} are the salient points in the query image that have verified correspondences in $I_g$.
Note that our use of one-to-one correspondences ensures that a visual element may have only a single correspondence in a given image.
As will be seen in detail below, the matching score $IniScore(c)$ allows us to incorporate our confidence concerning the reliability of the visual evidence contributed by individual visual elements into the overall $Score(g,q)$ used to rank the location.

Finally, we aggregate the visual elements and their scores per image in $I_g$ in order to generate the visual element cloud (cf. Fig.~\ref{fig:exam_LocVisClould}b).
Formally expressed, the visual element cloud $\mathbf{S}_g$ for location $g$ is calculated as:
\begin{equation}
\mathbf{S}_g=
\{ 
\mathbf{W}_{e}| e \in \mathbf{E}_g, \mathbf{W}_{e}=\{ w(e)_j | j=0,1...m(e) \} 
\}
\end{equation}
Here, $\mathbf{E}_g$ is the set of visual elements that occur in the query and link it with the images $I_g$ representing location $g$.
$\mathbf{W}_{e}$ is the set of weights $w(e)_j$ of correspondences between the visual element $e$ appearing in the query and the $j$\textsuperscript{th} image in $I_g$ in which it also appears.
The total number of images which have correspondences involving element $e$ in the set $I_g$ is denoted by $m(e)$.
 
\begin{figure}[h]
  \centering
  \scalebox{0.35}{\includegraphics[width=\textwidth]{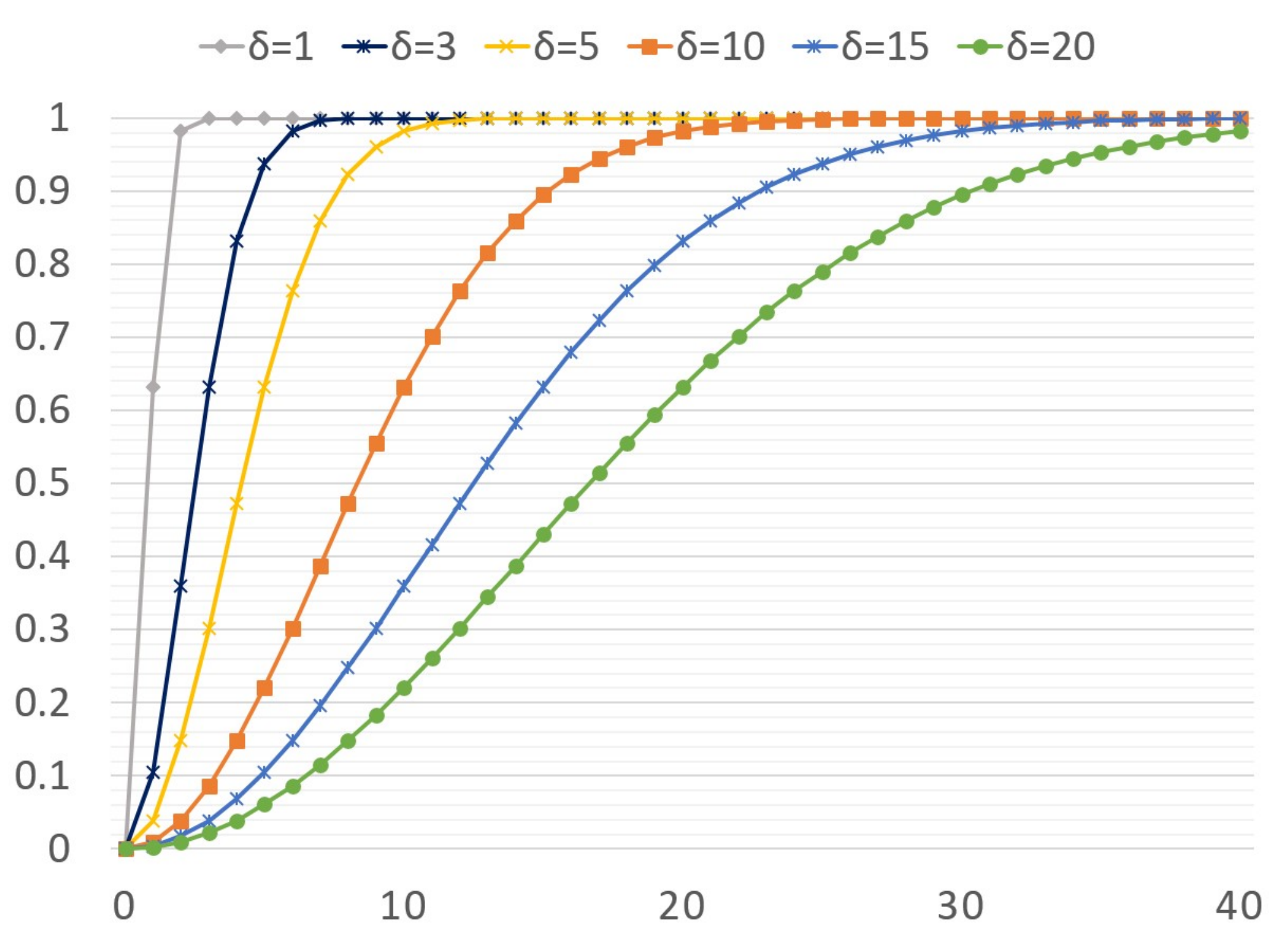}}
  \caption{Matching score smoothing function $w(e)_j$ vs. $IniScore(c)$ for various $\delta$.}\label{fig:PGMSmoothFun}
\end{figure}

The weights $w(e)_j$ are obtained by using a Gaussian function to smooth the initial matching score, $IniScore(c)$, of the correspondence $c$ in which the $j$\textsuperscript{th} appearance of the visual element $e$ is involved, and is denoted as
\begin{equation}
w(e)_j=1-\exp(-\frac{IniScore(c)^2}{\delta^2}).
\label{smoothFun}
\end{equation}
Here, $\delta$ controls the smoothing speed as shown in Fig.~\ref{fig:PGMSmoothFun}. 

The $\delta$ parameter is set according to the general, data-set independent, behavior of the geometric verification method that is employed. 
Note that when $\delta = 1$ the values of $w(e)_j$ are effectively either 0 or 1, meaning that visual elements either contribute or do not contribute, rather than being weighted.

\subsection{Geo-distinctiveness modeling}\label{SubSec_DVModel}
\begin{figure*}[t]
  \centering
  \scalebox{0.9}{\includegraphics[width=\textwidth]{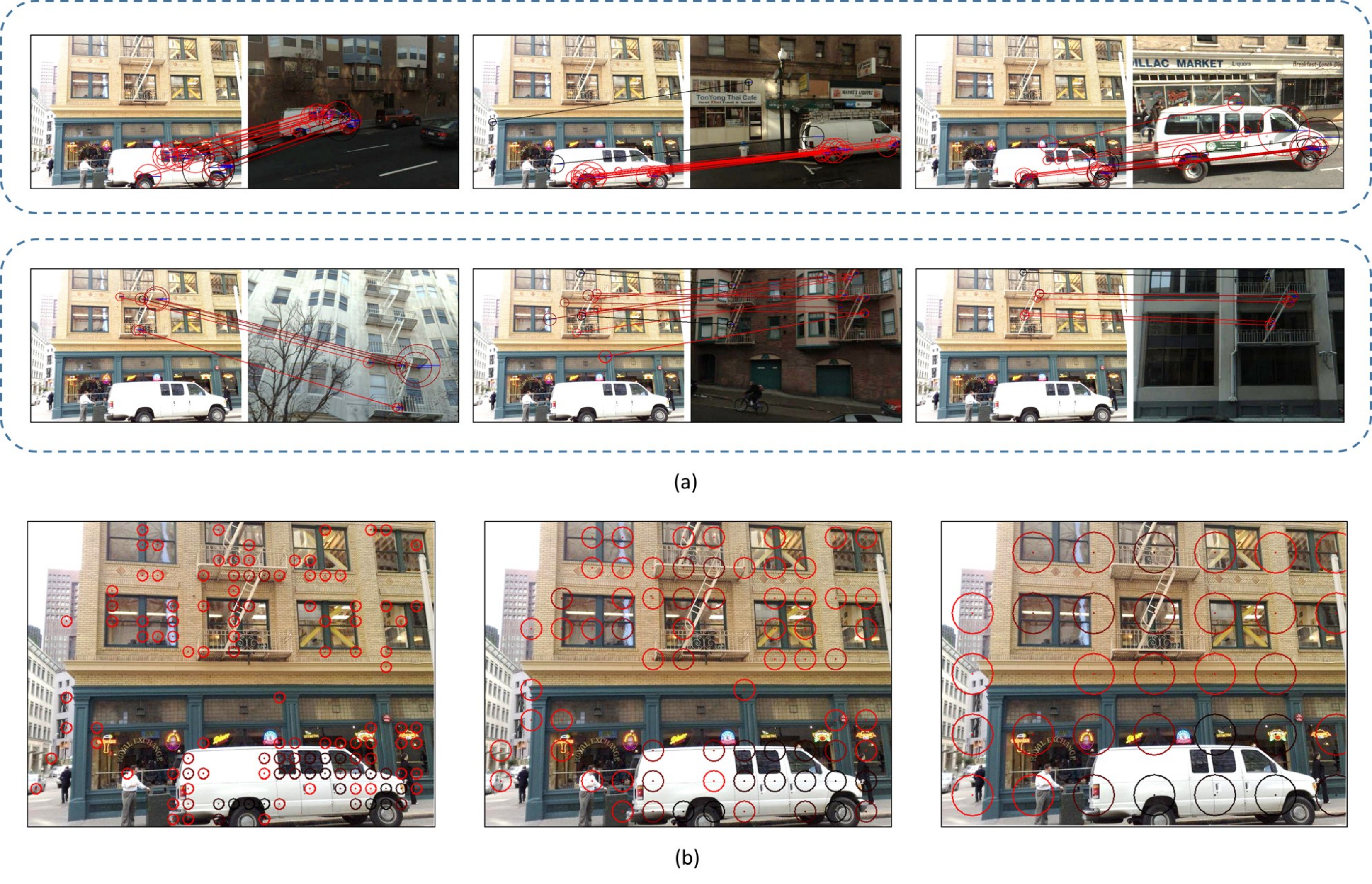}}
  \caption{Illustration of geo-distinctiveness modeling. Figure (a) shows how visual elements corresponding to two common objects in the query image (white delivery van and fire escape) give rise to strong matches with images from different locations. The geo-distinctiveness of these visual elements in the query image under different region resolution is shown in Figure (b), with the color changing from black to red to represent the increase of geo-distinctiveness.}\label{fig:exam_geoDistinctive}
\end{figure*}

We start our explanation of geo-distinctiveness modeling with an illustration of the basic mechanism.  
Fig.~\ref{fig:exam_geoDistinctive}(a) (top two rows) contain pairs of images. 
They show the correspondences between the query image (lefthand member of each pair) with images taken at locations other than the query location (righthand member of each pair).
As in the case of the visual element cloud, these correspondences pick out the visual elements that we use for further modeling.

Fig.~\ref{fig:exam_geoDistinctive}(b) (bottom row) shows how the geo-distinctiveness weights are calculated.
The image is divided into regions, and a geo-distinctiveness weight is calculated per region.
The three versions of the query image represent three different settings of region size, indicated by the increasing diameters of the circles.
In the figure, the center of the circle indicates the center of the region, and the color indicates the weight.
The color scale runs from red to black, with red indicating the most geo-distinctive regions.
Examination of Fig.~\ref{fig:exam_geoDistinctive}(b) shows the ability of geo-distinctiveness weights to focus in on specific, distinguishing features of the real world location.
Visual elements corresponding to common objects occurring at multiple locations (e.g., the white delivery van and fire escape) automatically receive less weight (i.e., as shown by black).
 
Expressed formally, geo-distinctiveness is calculated with the following process.
We divide the query image, of size $w \times h$, into non-overlapping small regions with size $\tilde{a} \times \tilde{a}$, $\tilde{a}=min(w/a,h/a)$.
For completeness note that we allow right and bottom regions to be smaller than $\tilde{a} \times \tilde{a}$, in the case that $w$ or $h$ is not an integer multiple of $a$.

We then transfer the scale represented by each visual element from the level of the neighborhood of a salient point to the level of an image region.
We carry out this transfer by mapping visual elements to the regions in which they are located.
Note that the consequence of this mapping is that all visual elements contained in the same image region are treated as the same visual element.
The effect of the mapping is to smooth the geo-distinctiveness of the visual elements in the query image. 
Changing $a$ will change the size of the region, and thereby also the smoothing.
The effect can be observed in Fig.~\ref{fig:exam_geoDistinctive}(b), e.g., the fire escape at the top middle of the photo is less discriminative (the circle turns black) as the area becomes larger.

For each visual element $e$ in each image in the image set $I_g$ for location $g$ in $G$, we calculate a geo-distinctiveness weight $W_{Geo}$. 
Recall that $e$ in each image in $I_g$ stands in a one-to-one correspondence $c$ with a visual element in the query image.
$W_{Geo}$ is then defined as 
\begin{equation}
W_{Geo}(e)=
\begin{cases}
	\log(N/n(r(e))),	& \text{if }n(r(e))< \vartheta \\
	0,  & \text{otherwise,}
\end{cases}
\label{IDF}
\end{equation}
where $N$ is the total number of location candidates (i.e., $|G|$),
$r(e)$ is the image region of the query containing the visual element corresponding to $e$, and
$n(r(e))$ is the total number of locations in $G$ with an image from their image set $I_g$ that is involved in a correspondence with any visual element occurring in the query region $r(e)$.
Finally, $\vartheta$ is a threshold completely eliminating the influence of elements that have correspondences with many locations in $G$.
The effect of parameters $a$ and $\vartheta$ is discussed in the experimental section.


\subsection{Visual matching per location}\label{SubSec_DVMatching}
\begin{figure}[t]
  \centering
  \scalebox{0.45}{\includegraphics[width=\textwidth]{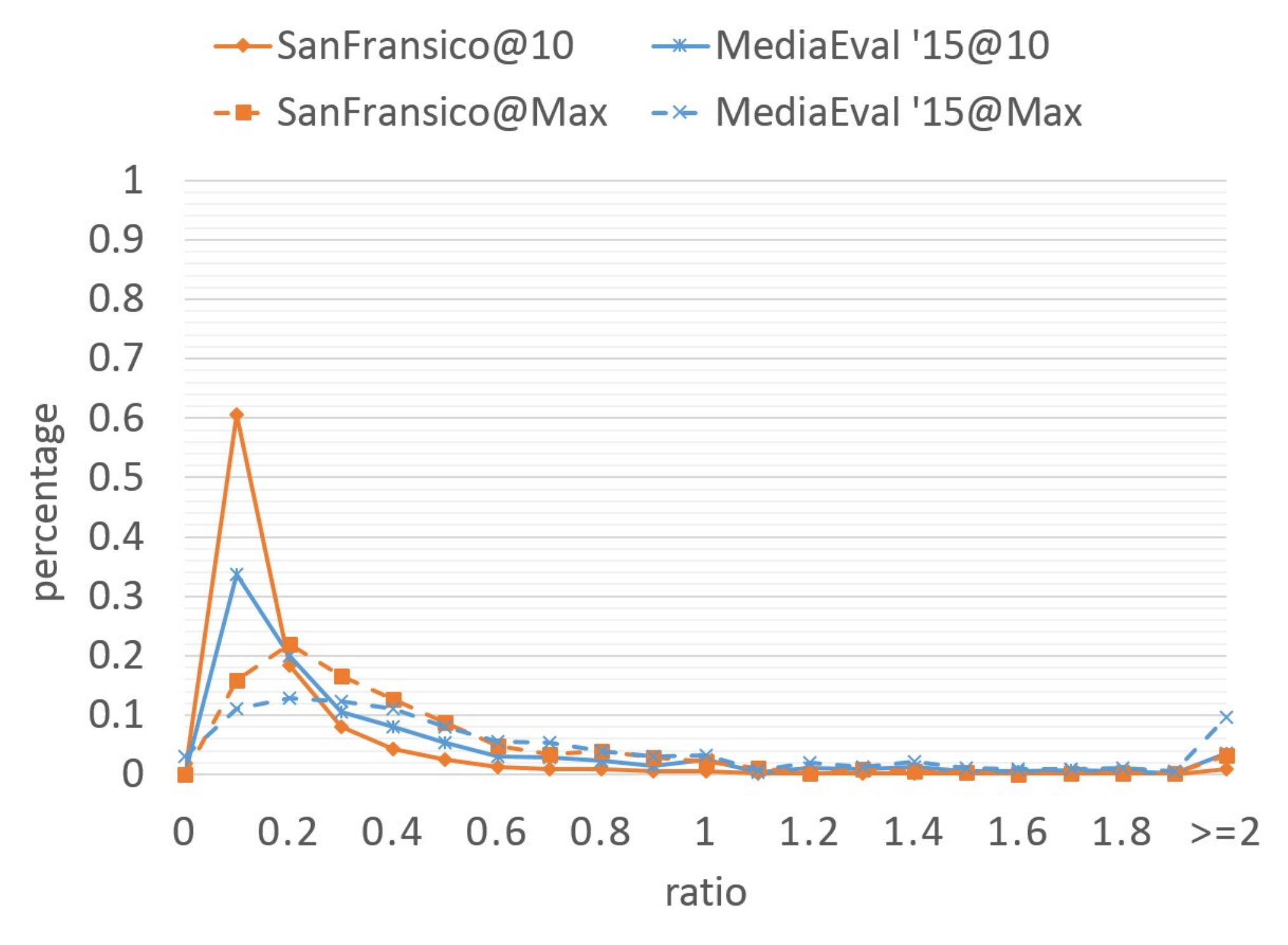}}
  \caption{Distribution of the ratio of number of unique visual elements between wrong location and true location. The scheme with @10 means the results are calculated based on the top-10 wrong locations in the initial ranked list for each query. The scheme with @Max means the results are calculated based on the wrong location that has the maximum number of visual elements among all wrong locations in the initial ranked list.}\label{fig:MatchNumDistribution}
\end{figure}

We start our discussion of visual matching by considering the patterns of visual elements associated with a true match between an query image and a location.
First, we investigate whether or not we can indeed expect more visual elements in true-location visual element clouds compared to wrong-location visual element clouds.
We carry out the analysis on two datasets, the San Francisco Landmark dataset and the MediaEval '15 Placing Task dataset, the geo-location estimation image sets used in our experiments, which will be described in detail in Section~\ref{ExperRes}.
Results are shown in Fig.~\ref{fig:MatchNumDistribution}. 
Here, we see that the ratio between the number of unique visual elements in a wrong-location cloud and a true-location cloud is mainly distributed between 0 and 1. 
The observation holds whether the top-10 ranked wrong locations are considered (solid line), or whether only the wrong location with the most visual elements is considered (dashed line).
This analysis points to the ability of the number of visual elements to distinguish true from wrong locations, and motivates us to include aggregation of visual elements as part of our visual matching model.

\begin{figure}[t]
  \centering
  \scalebox{0.45}{\includegraphics[width=\textwidth]{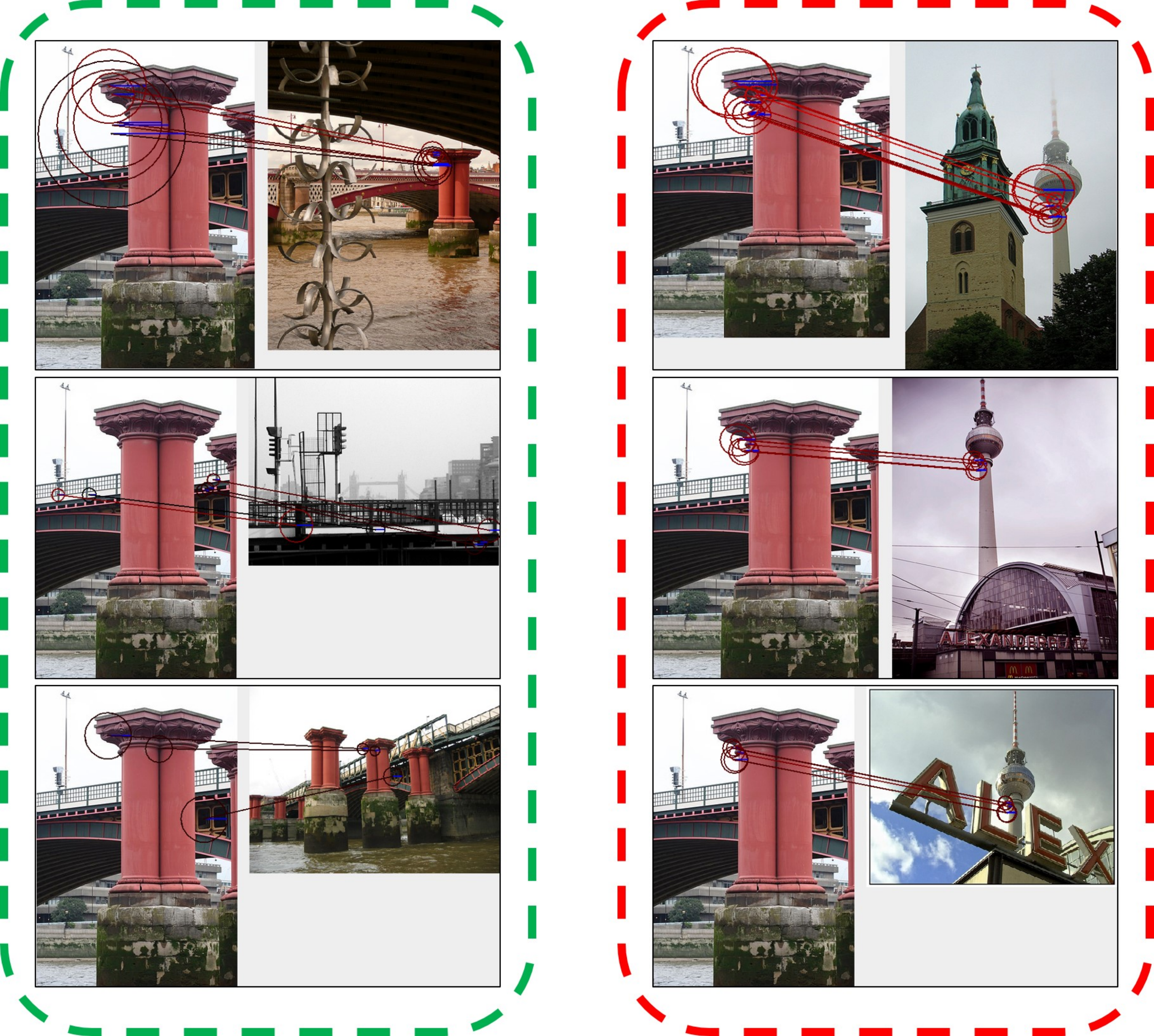}}
  \caption{Illustration of the initial correspondence set between the query image and the photos in two different locations with the color intensity from black to red representing the increase of the strength of the initial matching score. The left photo set is from the same location as the query image.}\label{fig:exam_locationMatching}
\end{figure}

Next, we return to our earlier statement (Section~\ref{Rationale}) that we expect the match between queries and a true location to display (a) a wider variety of visual elements, and (b) visual elements that are distributed over a greater area of the image, than in the case of a match with a false location. 
These expectations are borne out in our calculations of visual correspondences, as illustrated in Fig.~\ref{fig:exam_locationMatching}. 
The images from the true location (lefthand side) capture a broad and diverse view of the scene and thus match different regions of the query image, e.g., the column and the bridge, as opposed to the images taken at a wrong location (righthand side) that only have correspondences with few specific visual elements, e.g., the top of the column.  
This pattern leads us to not simply aggregate visual elements, but select them in a particular way.
Specifically, for a given area of the image query, only a single visual element is allowed to contribute per location. 
This approach rewards locations in which visual elements are diverse and distributed over the query image.

Expressed formally, visual matching uses the following procedure.
We divide the query, of size $w \times h$, into regions $\tilde{b} \times \tilde{b}$, $\tilde{b}=min(w/b,h/b)$. 
This splitting resembles what we used for geo-distinctiveness modeling, but serves a separate purpose in the current step. 
Then, in order to calculate the match between $q$ with a candidate location image set $I_g$, we iterate through each region of the query image.
For each region, we select the single visual element $e$ that has the strongest matching score with images from a given location. 
Recalling that $\mathbf{W}_{e}$ are the weights of the visual correspondences with the query for image set $I_g$ representing location $g$, the strongest matching score is expressed as $\tilde{w}_{e}=\max(\mathbf{W}_{e})$.
The result is a set of $k$ visual elements.
Note that although the same query image regions are used for all locations, $k$ may vary per location, and is less than the total number of query regions in the cases where some query regions fail to have links in terms of visual elements with a location.

The final visual proximity score between location $g$ and the query image $q$ combines a visual representation of the location $g$ and of the query $q$.
The representation of the query uses the visually distinctive weights $W_{Geo}(e)$ from Eq.~\ref{IDF}: 
$\mathbf{r}_{q}=(W_{Geo}(0),W_{Geo}(1),...,W_{Geo}(k))$. 
The representation of the location combines these weights with visual matching weights $\tilde{w}_{e}$: 
$\mathbf{r}_{g}=(\tilde{w}_{0}W_{Geo}(0),\tilde{w}_{1}W_{Geo}(1),...,\tilde{w}_{k}W_{Geo}(k))$.
The combination is calculated as,
\begin{equation}
Score(g,q)=\mathbf{r}_{q}\cdot{\mathbf{r}_{g}}=\sum\limits_{e\in\mathbf{E}_g} \tilde{w}_{e} W_{Geo}(e)^2
\end{equation}
The final location estimation for the query is calculated by ranking the locations by this score, and propagating the top-ranked location to the query.

\section{Experimental Setup}\label{ExperFrame}
In this section, we describe the setup of our experimental framework for assessing the performance of DVEM. This provides the background for our experimental results of parameter selection  (Section~\ref{sec:expar}), geo-contrained location estimation (Section~\ref{exp_SanFran}), geo-unconstrained location estimation (Section~\ref{exp_ME15}), and our comparison with the state of the art (Section~\ref{compStateOfArt}).

\subsection{Dataset}
We carry out experiments on two image datasets that are commonly used in location estimation, one for the geo-constrained, and one for the geo-unconstrained image geo-location prediction scenario.

\noindent\textbf{San Francisco Landmark dataset}~\cite{landmark_identification2011}: This dataset is designed for city-scale location estimation, i.e., geo-constrained location estimation. 
The database images (background collection) are taken by a vehicle-mounted camera moving around downtown San Francisco, and query images are taken randomly from a pedestrian's perspective at street level by various people using a variety of mobile photo-capturing devices. 
We use $1.06$M perspective central images (PCI) derived from panoramas as the database photos, and the original $803$ test images as queries.
For our detailed experiments in Sections~\ref{sec:expar} and~\ref{exp_SanFran} we use $10\%$ of the test images for development, and report results on the other $90\%$ of the test images. 
The ground truth for this dataset consists of building IDs. The geo-location of an image is considered correctly predicted if the building ID is correctly predicted.

\noindent\textbf{MediaEval '15 Placing Task dataset}~\cite{PlacingWorkNote2015}: This dataset is designed for global scale location estimation, i.e., geo-unconstrained location estimation. 
It is a subset of the YFCC100M collection~\cite{fcc100m2015}, a set of Creative Commons images from Flickr, an online image sharing platform.
The background collection and the query images were randomly selected in a way that maintained the global geographic distribution within the online image sharing community. 
The MediaEval 2015 Placing Task dataset is divided into $4.6$M training and $1$M test images.
Here again for our detailed experiments in Sections~\ref{sec:expar} and~\ref{exp_ME15} we use $2\%$ of the test set for development, and report results on the other $98\%$ of the test set.
The ground truth for this dataset consists of geo-coordinates, either recorded by the GPS of the capture device or assigned by hand by the uploading users.
An image is considered to be correctly predicted if its predicted geo-coordinates fall within a given radius $r_{eval}$ of the ground truth location. 
$r_{eval}$ controls the evaluation precision and the tolerance of the evaluation to noise in the ground truth.

\subsection{Computing visual similarity}
Conceptually, we consider the visual matches 
between different areas 
of two images as evidence that their visual content reflects the same location in the physical world, possibly differing as to how they are captured, e.g., capturing angle, scale or illumination. 
In order to identify these
areas
and assess the strength of the link between their occurrences in images, we deploy our recently-developed image retrieval system~\cite{my_CVPR15_PGM}. 
This system is based on pairwise geometric matching technology and is built upon the standard bag-of-visual-words paradigm. 
The paradigm is known to scale up well to a large-scale datasets~\cite{videoGoogle,burstiness_CVPR09,ThreeThings_CVPR12}. 
To further speed up retrieval and improve accuracy, we use pairwise geometric matching in the following pipeline of state-of-the-art solutions:
\begin{itemize}
	\item Features \& Vocabularies: Since up-right Hessian-Affine detector and Root-SIFT~\cite{ThreeThings_CVPR12} have proven to yield superior performance, we use this feature setting to find and describe invariant regions in the image. We use exact k-means to build the specific visual world vocabularies with the size of 65,536 based on the features from the training images.
	\item Multiple Assignment: To address the quantization noise introduced by visual word assignment, we adopt the strategy used in~\cite{HE_J2010}, which assigns a given descriptor to several of the nearest visual words. As this multiple assignment strategy significantly increases the number of visual words per image, we only apply this at the query side. 
	\item Initial ranking: We adopt the Hamming Embedding technique combined with burstiness weighting proposed in~\cite{burstiness_CVPR09} in the initial ranking phase.
	\item Geometric verification: To find the reliable correspondences for DVEM, the pairwise geometric matching technology~\cite{my_CVPR15_PGM} is employed for fast geometric verification, which is reported to be the state-of-the-art in image retrieval in terms of speed and accuracy. In the experiment conducted on the development set, we found that due to a high inter-similarity of the street view images taken in downtown San Francisco, removing the correspondences with low matching score generated by pairwise geometric matching can generally help to improve the estimation. Here the threshold is set to $4$. 
\end{itemize}
The ranked list resulting from this computation of visual similarity is used in the Candidate Image Selection step (cf. Fig.~\ref{fig:SysScheme}) and for two baselines, as discussed next.

\subsection{Experimental design}
We carry out two different sets of evaluations that compare the performance of DVEM to the leading content-based approaches to image geo-location estimation.
The first set (Sections~\ref{exp_SanFran} and~\ref{exp_ME15}) assesses the ability of DVEM to outperform other search-based geo-location estimation approaches, represented by VisNN and GVR:
\begin{itemize}
  \item \emph{VisNN}: Our implementation of the 1-NN approach~\cite{IM2GPS}, which uses the location of the image visually most similar to the query image as the predicted location. It is a simple approach, but in practice has proven difficult to beat.
  \item \emph{GVR}: Method used in~\cite{my_TMM_GVR}, which expands the candidate images by their locations and uses the overall visual similarity of images located in one location as the ranking score for that location. 
  This method is chosen for comparison since it has been demonstrated to outperform other state-of-the-art approaches for geo-unconstrained location estimation~\cite{my_GVR_ICMR13,my_MedEvl13}.
\end{itemize}
The second set of evaluations (Section~\ref{compStateOfArt}) compares our methods with other state-of-art methods, which do not necessarily use a search-based framework. 

Our evaluation metric is Hit Rate at top $K$ ($HR@K$). 
Recall that given a query, the system returns a ranked list of possible locations. 
$HR@K$ measures the proportion of queries that are correctly located in the top $K$ listed locations. 
Specifically, $HR@1$ represents the ability of the system to output a single accurate estimate.

\section{Experimental results} \label{ExperRes}
We implemented our DVEM framework on top of the object-based image retrieval system~\cite{my_CVPR15_PGM} by constructing a Map-Reduce-based structure on a Hadoop-based cluster\footnote{This work was carried out on the Dutch national e-infrastructure with the support of SURF Foundation.} containing 1,500 cores. 
The initial visual ranking (the candidate image selection step) takes about $105$ mins for San Francisco dataset ($803$ queries on a collection of $1.06$M photos) and about $88$ hours for the MediaEval '15 dataset ($1$M queries on a collection of $4.6$M photos). 
The DVEM stage is executed after the initial visual ranking, and takes $2.4$ms per query.

In this section, we report the experimental results and compare our DVEM method with reference methods in both areas of geo-constrained and geo-unconstrained location estimation.
We use part of the test data ($10 \%$ for San Francisco dataset and $2 \%$ for MediaEval '15 dataset) as development partition to set the parameters of DVEM, and use the rest of the test data to evaluate the system. 
Recall that the parameters are the image region size $a$ defined in Section~\ref{SubSec_DVModel}, the frequent threshold $\vartheta$ defined in Eq.~\eqref{IDF} and the image region size $b$ defined in Section~\ref{SubSec_DVMatching}. 
The parameter $\delta$ defined in Eq.~\eqref{smoothFun} is set empirically to $5$ based on the general observation that the initial correspondence score generated by pairwise geometric matching~\cite{my_CVPR15_PGM} usually reflects a strong match when it is above $10$. 
As previously mentioned, the number of top-ranked images from the image retrieval system, which are used to generate the candidate locations set $G$, is set to $1000$. Note that we use the same $G$ for GVR.

\subsection{Impact of the parameters}
\label{sec:expar}
We start our series of experiments by evaluating the impact of $a$, $b$,  $\vartheta$ on the system performance using our development partition. 
We explore the parameter space with grid search, as shown in Table~\ref{tab:Para}.
For both $a$ and $b$, we considered the values 0, 30, 20 and 10 (Table~\ref{tab:Para}, top).
Note that $a=0$ means that the system assigns a different geo-distinctiveness weight to each individual visual element, and $a=30,20,10$ are regions increasing in size. 
Similarly, $b=0$ means that system deploys all visual elements appearing in the images of a given location for query-location matching,  and $b=30,20,10$ are regions increasing in size.
After 10 performance dropped dramatically, and these values were not included in the table.
We choose $a=10, b=20$ as an operating point for the San Francisco dataset and $a=0, b=30$ for the MediaEval '15 dataset.
For $\vartheta$, we considered the values 4, 5, 6 and 7, but found little impact (Table~\ref{tab:Para}, bottom).
We choose $\vartheta=5$ for the San Francisco dataset and $\vartheta=6$ for the MediaEval dataset.

\begin{figure}[t]
  \centering
  \scalebox{0.45}{\includegraphics[width=\textwidth]{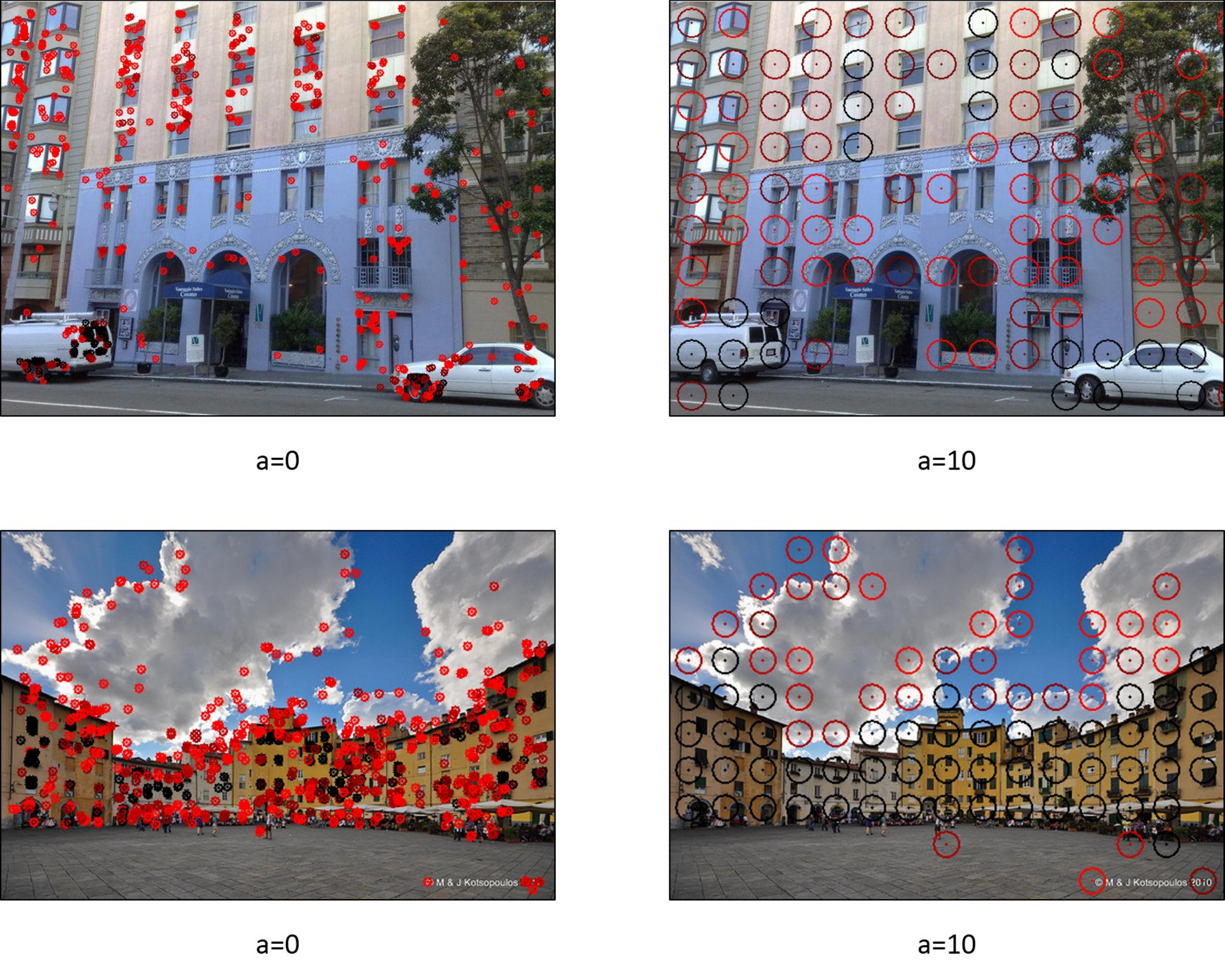}}
  \caption{Illustration of the geo-distinctiveness of visual elements under different region resolutions for query images from San Francisco dataset (top row) and MediaEval '15 dataset (bottom row). The color changing from black to red indicates an increase in geo-distinctiveness}\label{fig:ana_Parameters}
\end{figure}

\begin{table}[htb!]\renewcommand{\arraystretch}{1.3}
\caption{HR@1(\%) comparison of DVEM on San Francisco (fixed ground truth) and MediaEval '15 datasets ($r_{eval}=1km$) with different $a$, $b$, and $\vartheta$.}\label{tab:Para}
\vspace{-0.5em}
\centering
\begin{tabular}{c | c c c c | c c c c}

\multicolumn{1}{c}{} & \multicolumn{4}{c}{San Francisco}       				& \multicolumn{4}{c}{MediaEval '15}				\\

\multicolumn{1}{c}{} & \multicolumn{4}{c}{$\vartheta=5$}       				& \multicolumn{4}{c}{$\vartheta=6$}				\\
\hline
\backslashbox{a}{b} & 0 	& 30 	& 20	& 10 							& 0 	& 30 	& 20	& 10 \\
\hline
0					& 81.3	& 80	& 82.5	& 81.3							& 8.1 	& \textbf{8.2} 	& 8.1	& 8 \\
\hline
30					& 80	& 80	& 81.3	& 80							& 8 	& 7.9 	& 7.9	& 7.8 \\
\hline
20					& 81.3	& 81.3	& 80	& 80							& 7.8 	& 7.8 	& 7.8	& 7.6 \\
\hline
10					& 80	& 82.5	& \textbf{83.8}	& \textbf{83.8}			& 7.2 	& 7.3 	& 7.3	& 7.2 \\
\hline

\multicolumn{9}{c}{}\\
\multicolumn{1}{c}{} & \multicolumn{4}{c}{$a=10,b=20$}       					& \multicolumn{4}{c}{$a=0,b=30$}				\\
\hline
$\vartheta$ 			 & 4 	& 5 	& 6		& 7 								& 4 	& 5 	& 6		& 7 \\
\hline
					 & \textbf{83.8} & \textbf{83.8} & \textbf{83.8} & 82.5		& 8.1 	& 8.1 	& \textbf{8.2}	& 8.1 \\
\hline
\end{tabular}
\end{table}

We notice that the performance is mainly influenced by the parameter $a$, which is used to smooth the geo-distinctiveness of the visual elements in the query. 
The optimal values for parameter $a$ are different on the two datasets. 
An investigation of the difference revealed that it can be attributed to the difference in the respective capture conditions.
The examples in Fig.~\ref{fig:ana_Parameters} illustrate the contrast.
The queries in the San Francisco dataset (Fig.~\ref{fig:ana_Parameters}, top) are typically zoomed-in images, taken on the street with a limited distance between the camera and the captured object (e.g., car or building). 
High levels of zoom results in the salient points that correspond to object details, e.g., a single tire on a car can have multiple salient points assigned to it. 
Such a high resolution of salient points may confuse object matching and is for this reason not productive for location estimation. 
For this reason, it appears logical that a value of $a$ that leads to a higher level of grouping of salient points for the purposes of geo-distinctiveness assessment leads to the best performance.
In contrast, the queries in the MediaEval '15 dataset that have the best potential to be geo-located (Fig.~\ref{fig:ana_Parameters}, bottom) are mostly zoomed-out images capturing a scene from a distance. The level of detail is much less than in the previous case, and the salient points tend to already pick out object-level image areas relevant for location estimation. Aggregating the salient points together through image splitting like in the previous case would have a negative effect, as it would reduce the resolution of salient points too drastically, leading to a loss of geo-relevant information. 
For this reason, it is logical that the parameter value $a=0$ is the optimal one, reflecting that no image splitting should be carried out. 

\subsection{Geo-constrained location estimation}\label{exp_SanFran}

\begin{figure}[h]
  \centering
  \subfigure[] {\scalebox{0.24}{\includegraphics[width=\textwidth]{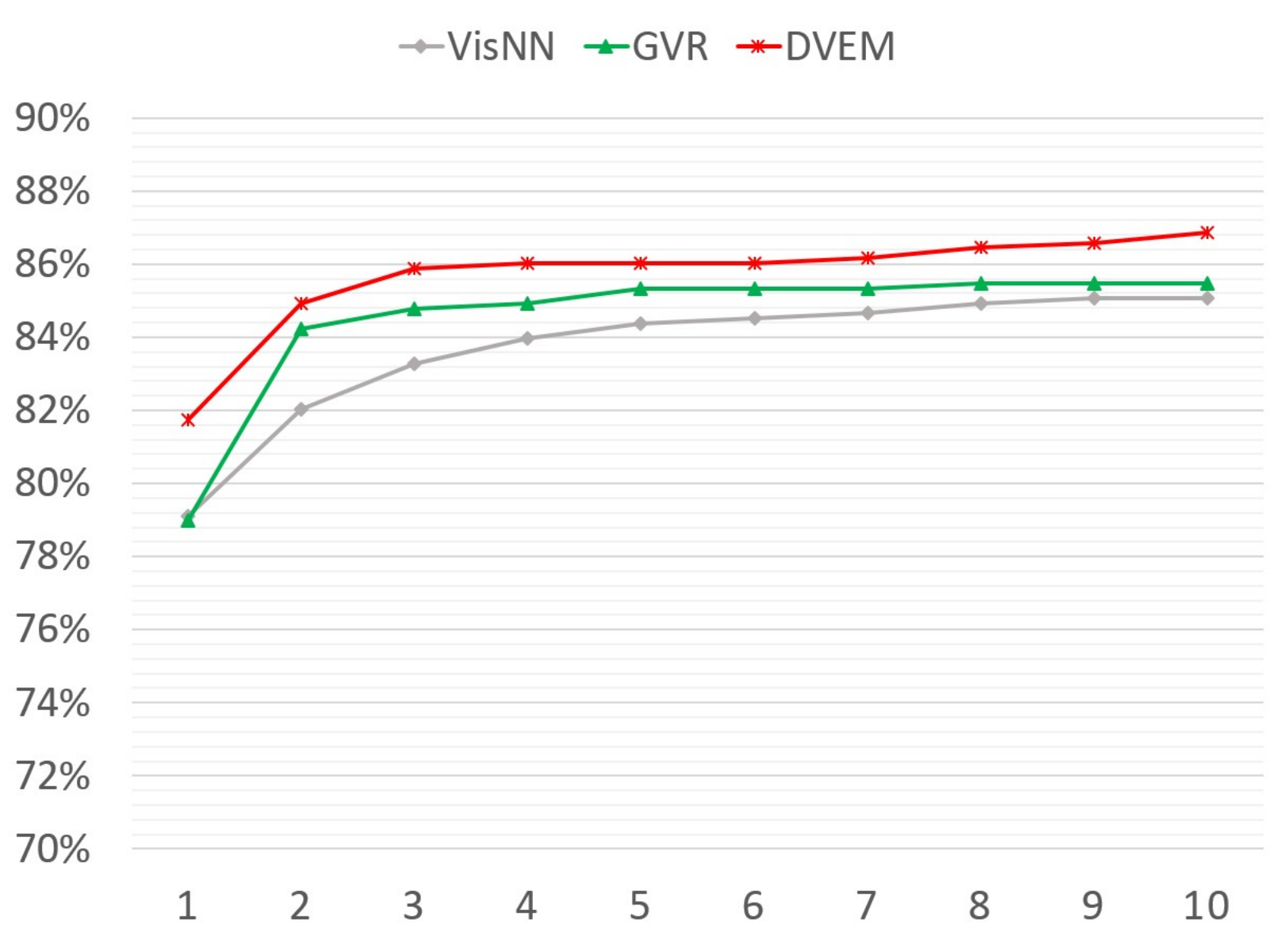}}}
  \subfigure[] {\scalebox{0.24}{\includegraphics[width=\textwidth]{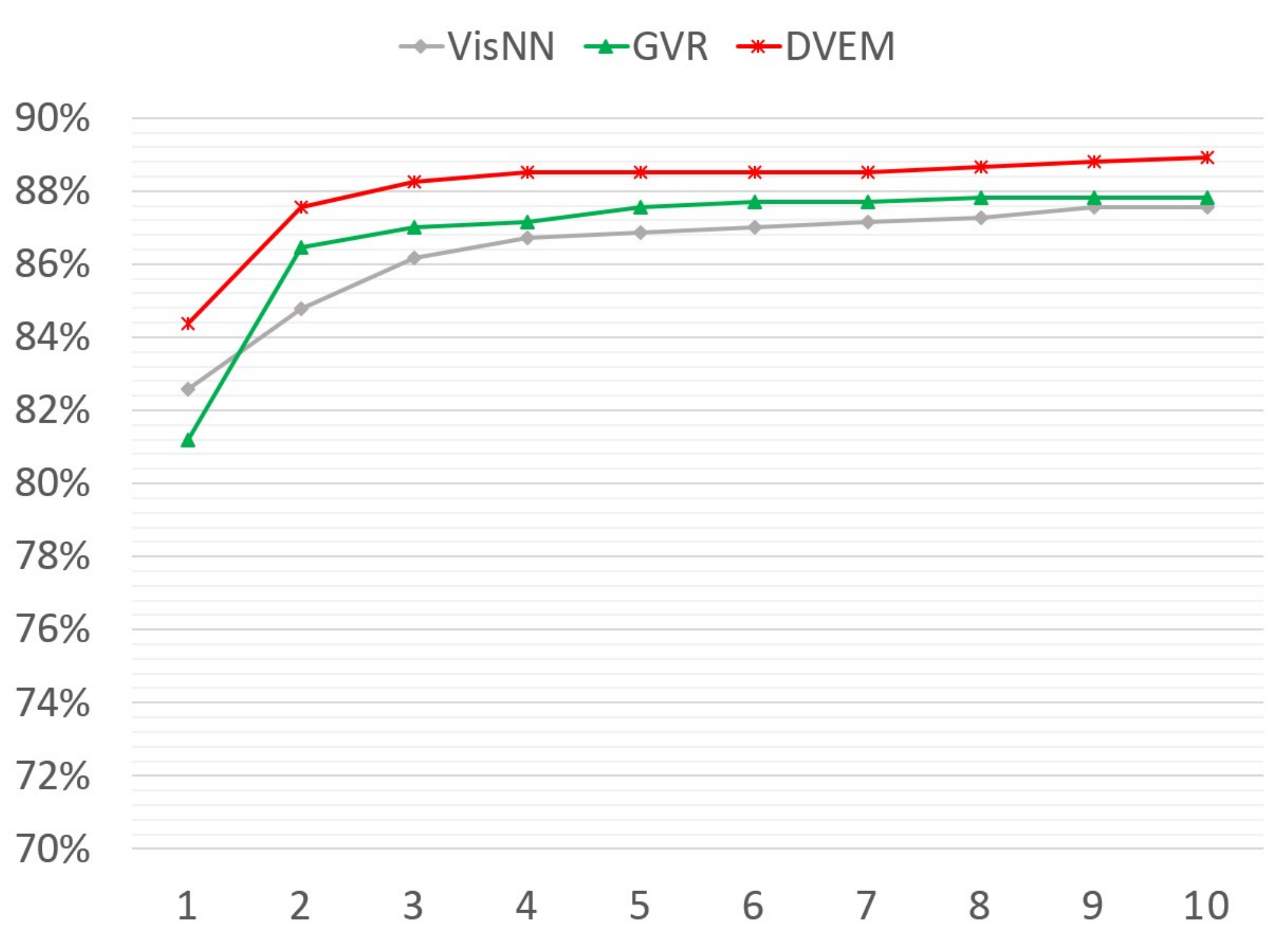}}}
  \vspace{-0.5em}
  \caption{HR@k performance for varying $k$ on the SanFrancisco street view dataset. (a) performance with respect to the original ground truth, (b) performance with respect to the fixed ground truth released on April 2014.}\label{fig:res_SanFran}
\end{figure}

The performance of different methods on the San Francisco Landmark dataset is illustrated in Fig.~\ref{fig:res_SanFran}. DVEM consistently outperforms both VisNN and GVR across the board, with the performance gain of $3\%$ and $4\%$ for HR@1 with respect to the fixed ground truth released in April 2014 (Fig.~\ref{fig:res_SanFran}.b). 

GVR performs even worse than VisNN with respect to the fixed ground truth. This is due to the fact that in the street-view dataset the database images are captured by the survey vehicle, which can make multiple near-duplicate images per location. When a location contains same visual elements of the query image, e.g., the white van in Fig.~\ref{fig:exam_geoDistinctive}b, the summed visual similarity of images taken in this location will heavily influence the estimation. 
In contrast, DVEM can handle this situation since it differentiates visual elements based on their geo-distinctiveness and eliminates the influence of redundancy by matching not at the image level, but rather at the level of the visual element cloud. 
  
We note that, as $52$ out of $803$ ($6.5\%$) query images do not correspond in location to any images in the database collection.
Consequently, the maximal performance that can be reached is $93.5\%$. 
In addition, the ground truth is automatically labeled based on building ID, which is generated by aligning images to a 3D model of the city consisting of $14$k buildings based on the location of the camera~\cite{landmark_identification2011}. 
This introduces noise into the ground truth. 
We conducted a manual failure analysis on the $74$ queries for which DVEM makes wrong estimation with respect to HR@1. 
We found that for $9$ queries, the ground-truth database images are irrelevant, and for $32$ queries, the database images located in the top-1 predicted location are relevant, but their building ID is not included in the ground truth. 
This makes the maximum performance that could be achieved by DVEM an HR@1 of $88.3\%$.


\subsection{Geo-unconstrained location estimation}\label{exp_ME15}
\begin{figure}[t]
  \centering
  \subfigure[] {\scalebox{0.24}{\includegraphics[width=\textwidth]{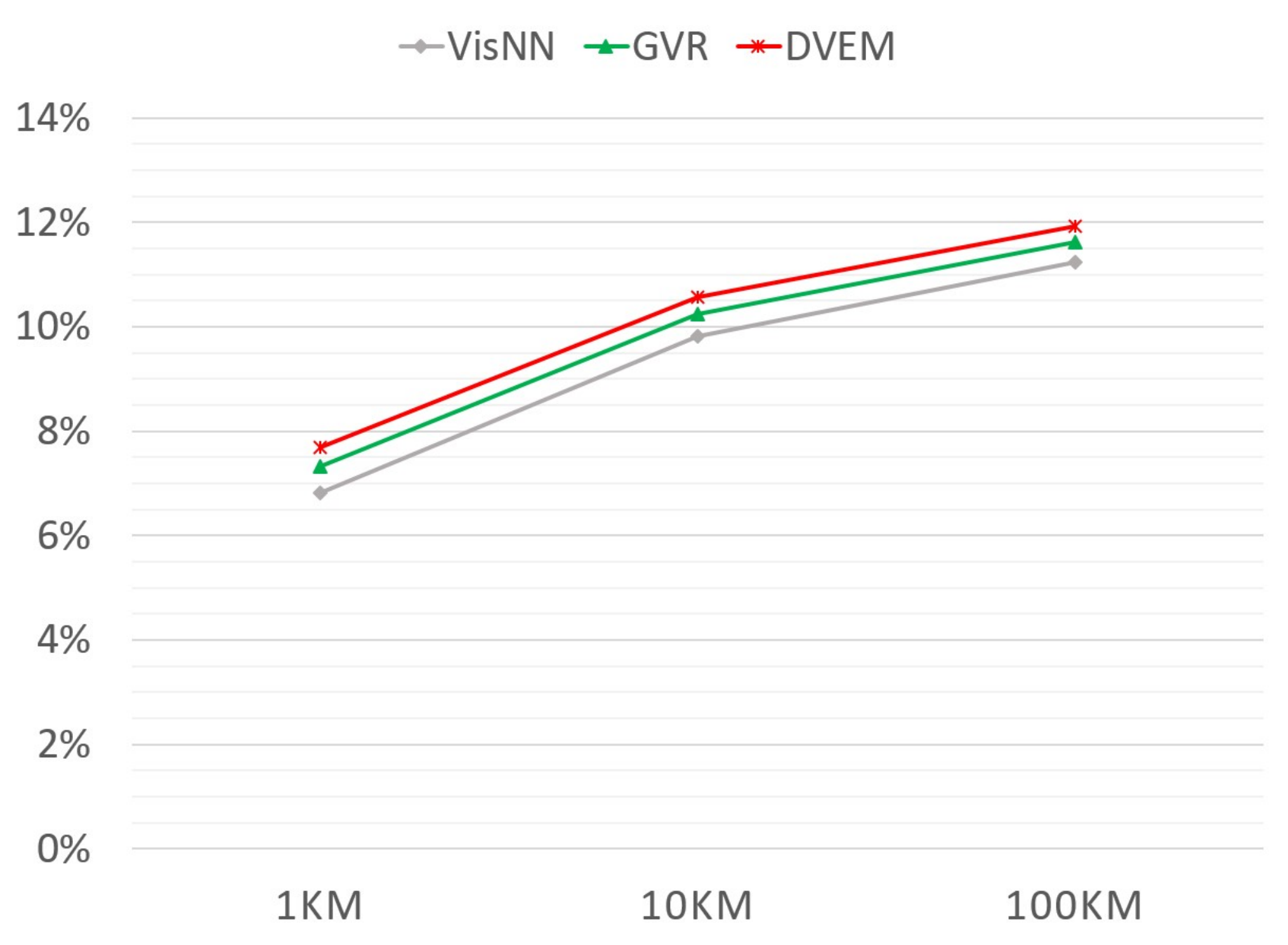}}}
  \subfigure[] {\scalebox{0.24}{\includegraphics[width=\textwidth]{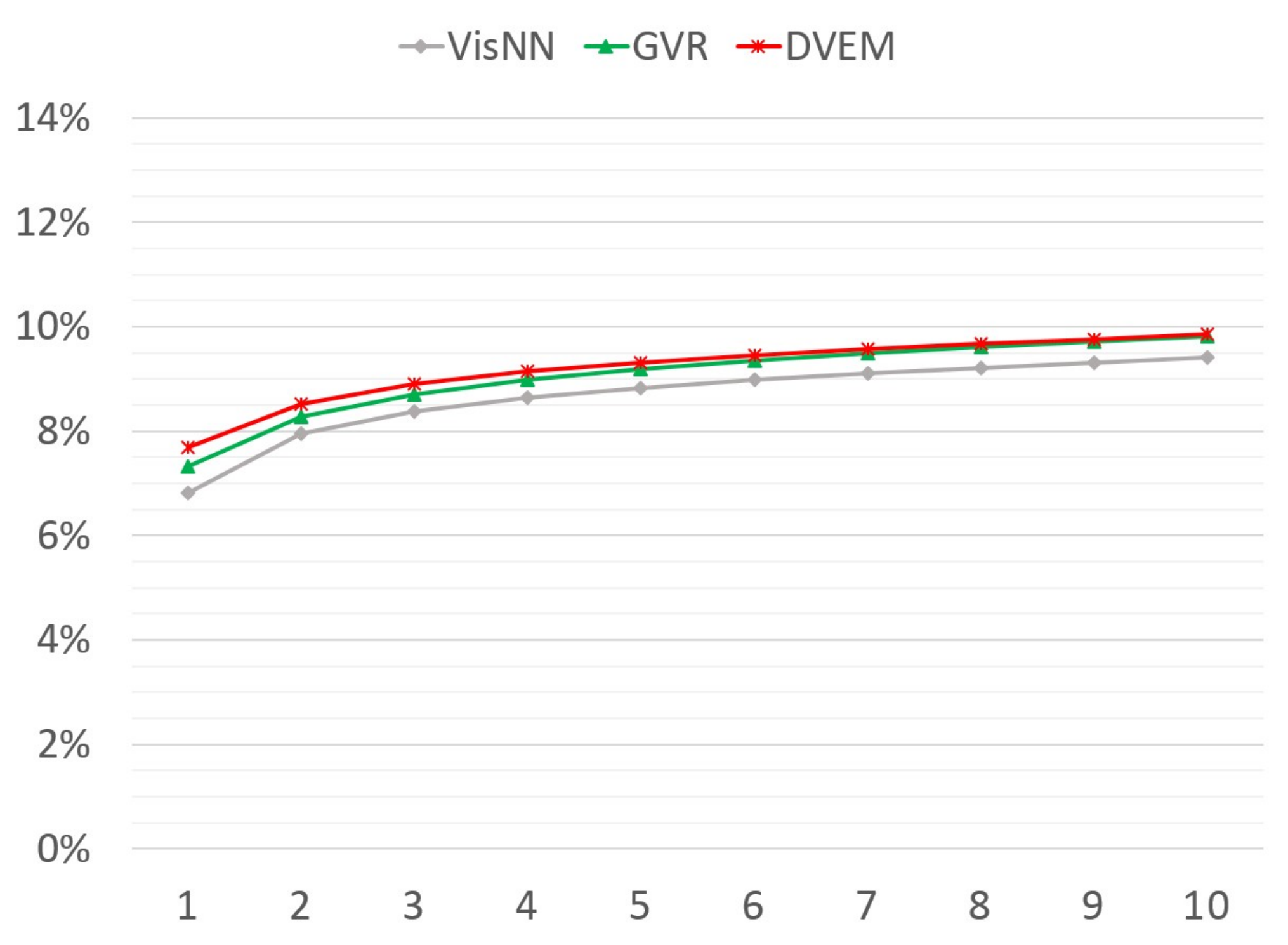}}}
  \vspace{-0.5em}
  \caption{Performance on the MediaEval '15 Placing Task dataset. (a) HR@1 with respect to different evaluation radiuses, (b) HR@k performance for varying $k$ and for the evaluation radius of $1km$.}\label{fig:res_ME15}
\end{figure}

Fig.~\ref{fig:res_ME15} shows the performance of different methods with different values of $r_{eval}$ (Fig.~\ref{fig:res_ME15}a.) and different hit rates (Fig.~\ref{fig:res_ME15}b.) on the MediaEval '15 Placing Task dataset. 
This figure demonstrates that DVEM consistently outperforms both VisNN and GVR. 
The gain in performance is $12\%$ over VisNN and $5\%$ over GVR for HR@1. 

\begin{figure*}[t]
  \centering
  \scalebox{1}{\includegraphics[width=\textwidth]{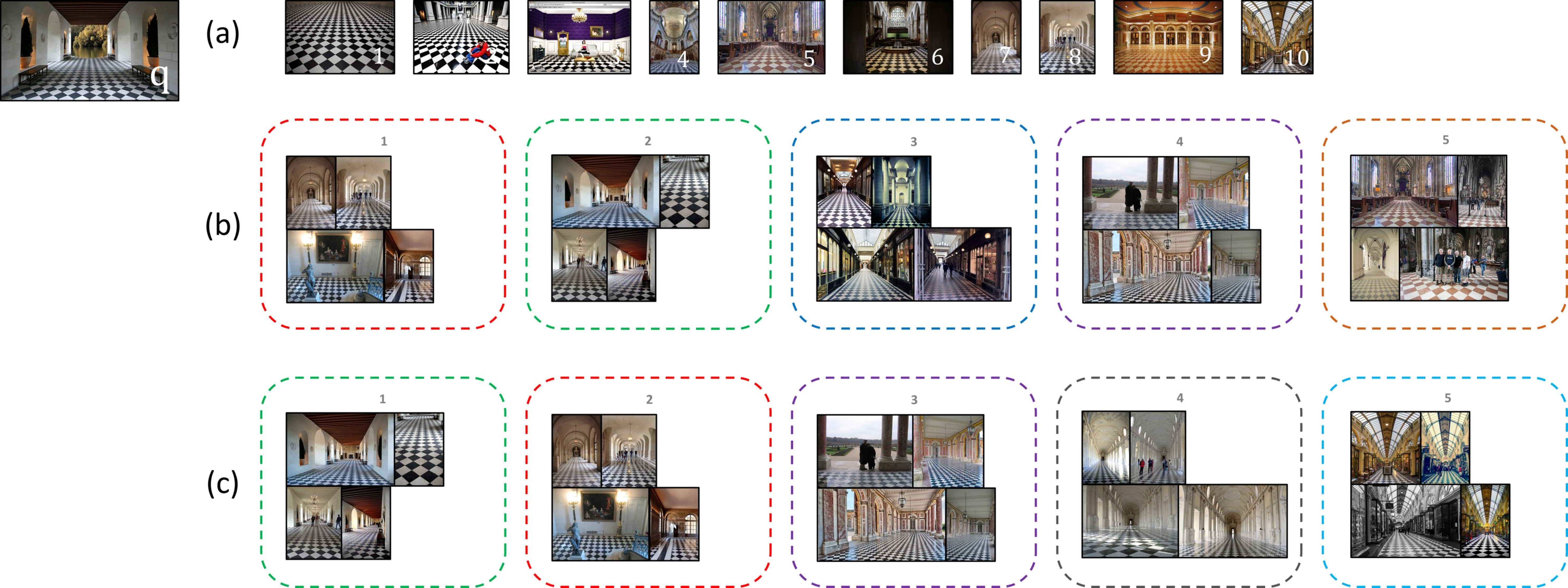}}
  \vspace{-0.8em}
  \caption{Illustration of the relative performance among the methods VisNN, GVR and DVEM on the MediaEval '15 Placing Task dataset: (a) the initial visual rank of top-10 most similar photos for a given query, the location of the top ranked photo is the result of VisNN, (b) ranked candidate locations using GVR, (c) ranked candidate locations using DVEM. There are maximum $4$ photos shown for each location.}\label{fig:showRankRes_ME15}
\end{figure*}

Next we turn to investigate in more detail why VisNN is outperformed by GVR, which is in turn outperformed by our new DVEM approach.
In general, GVR outperforms VisNN because it can leverage the existence of multiple images from the true location that are visually similar to the query.
GVR fails, however, when wrong locations also are associated with multiple images that are visually similar to the query. 
DVEM, however, is able to maintain robust performance in such cases.
Fig.~\ref{fig:showRankRes_ME15} contains an example that illustrates the difference.
The query $q$ is shown on the left. 
VisNN is illustrated by row (a), which contains the top-10 images returned by VisNN.
There is no correct image for the query location among them.
This reflects that the collection lacks a single good image-level visual match for the query.
GVR is illustrated by row (b), which contains five sets of images from the five top-ranked candidate locations.
We see the top-1 candidate location image set contains many images similar to the query, although it is not the true location.
Instead, the true location, whose candidate location image set also contains many images, is ranked second.
DVEM is illustrated by row (c), which again contains five candidate location image sets.
This time, the correct location is ranked first.
We can see that the DVEM decision avoided relying too heavily on the distinctive floor pattern, which is common at many tourist locations, and cause GVR to make a wrong prediction.
Instead DVEM is able to leverage similarity matches involving diverse and distributed image areas (such as the ceiling and the alcoves in the walls), favoring this evidence over the floor, which is less geo-distinctive.

\subsection{Comparison with the state-of-the-art}\label{compStateOfArt}
In this experiment, we compare DVEM with other state-of-the-art location estimation systems regarding both the geo-constrained and geo-unconstrained case. 
We compare our results with the top results that have been reported by other authors on the two experimental datasets that we use. 

\begin{figure}[t]
  \centering
  \subfigure[] {\scalebox{0.24}{\includegraphics[width=\textwidth]{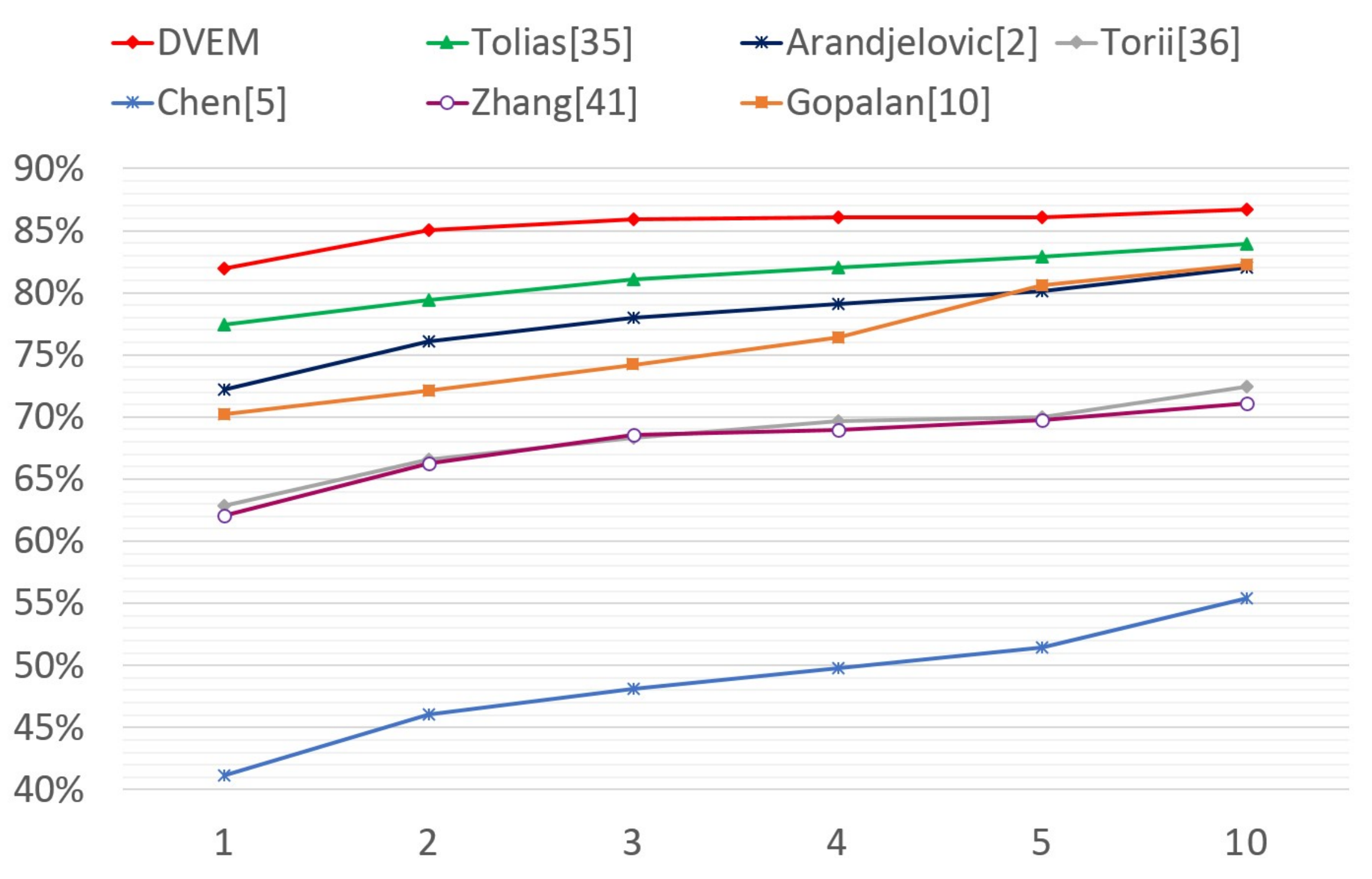}}}
  \subfigure[] {\scalebox{0.24}{\includegraphics[width=\textwidth]{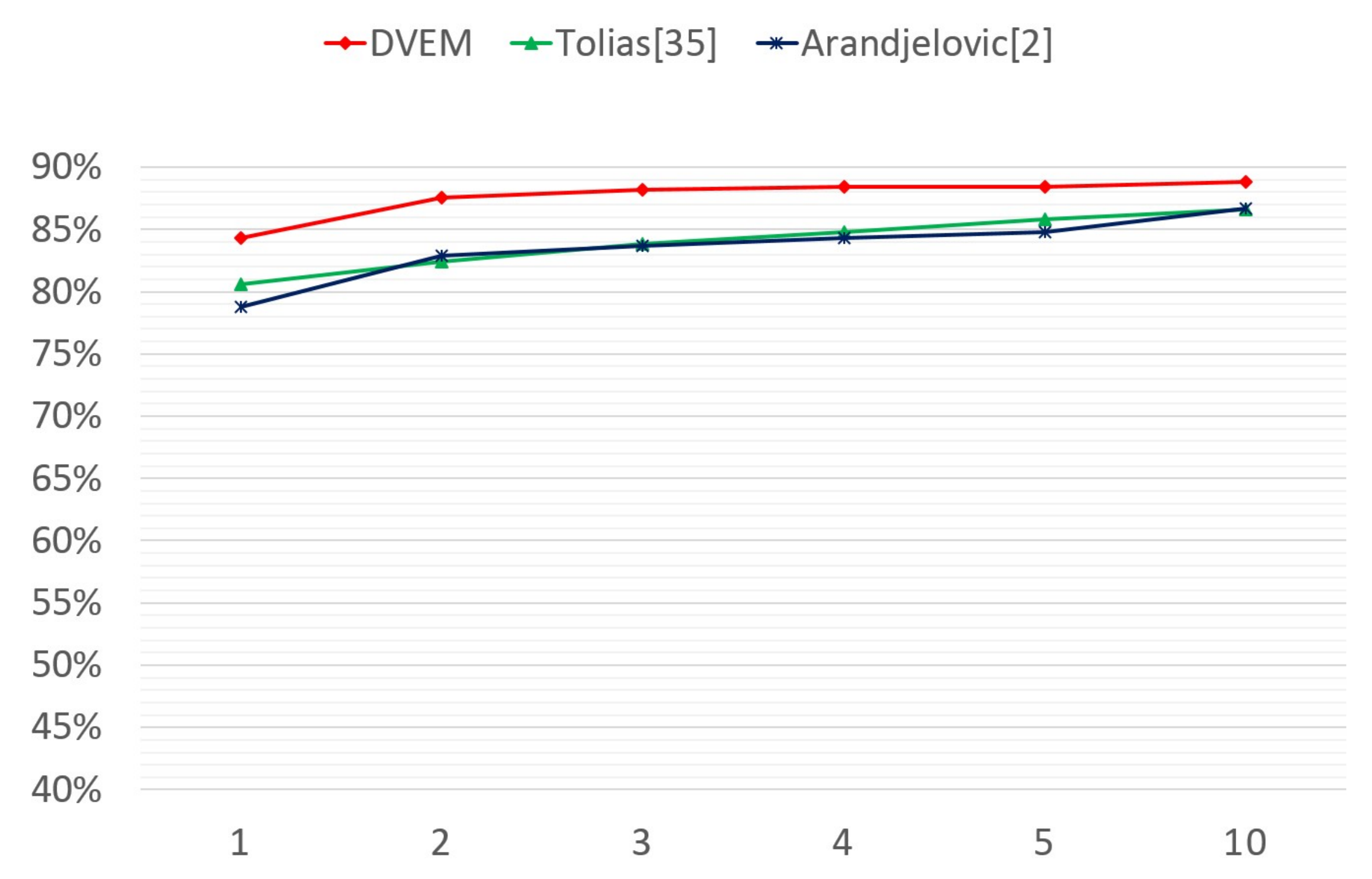}}}
  \vspace{-0.5em}
  \caption{HR@k performance for varying $k$ on the SanFransico street view dataset. (a) performance with respect to the original ground truth, (b) performance with respect to the fixed ground truth released on April 2014.}\label{fig:res_SanFran_stateOfArt}
\end{figure}

As the reference methods for geo-constrained location estimation, we use the work of Gopalan (2015)~\cite{SpareseCodingGeoLoc_CVPR15}, Tolias et al. (2015)~\cite{ASMK_IJCV15}, Arandjelovi\'{c} and Zisserman (2014)~\cite{DisLocation2014}, Torii et al. (2013)~\cite{repetStru_CVPR13}, Zhang et al. (2012)~\cite{zhang2012Qfusion} and the initial work of Chen et al. (2011)~\cite{landmark_identification2011}. 
Results are reported on the entire test set as defined with the San Francisco dataset release.
This set is identical to the sets on which these authors report their results.
The results in Fig.~\ref{fig:res_SanFran_stateOfArt} demonstrate that our proposed DVEM approach outperforms the state-of-the-art on the \emph{San Francisco} dataset.
For completeness, we include additional discussion of our experimental design.
The papers cited in Fig.~\ref{fig:res_SanFran_stateOfArt} use a variety of tuning methods, which are sometimes not fully specified.
We assume that these tuning methods are comparable to our choice, namely to use, 10\% of the test data (Section~\ref{sec:expar}).
Referring back to Table~\ref{tab:Para}, we can see that our demonstration of the superiority of DVEM is independent of this assumption.
In the table, we see that the difference in performance for DVEM for the best and the worst parameter settings is less than 4\% absolute.
If the performance of a very poorly tuned version of DVEM falls by this amount, it still remains competitive with well-tuned versions of the other approaches in Fig.~\ref{fig:res_SanFran_stateOfArt}.
This assures us that the superiority of our approach does not lie in our choice of tuning.

\begin{figure}[t]
  \centering
  \scalebox{0.4}{\includegraphics[width=\textwidth]{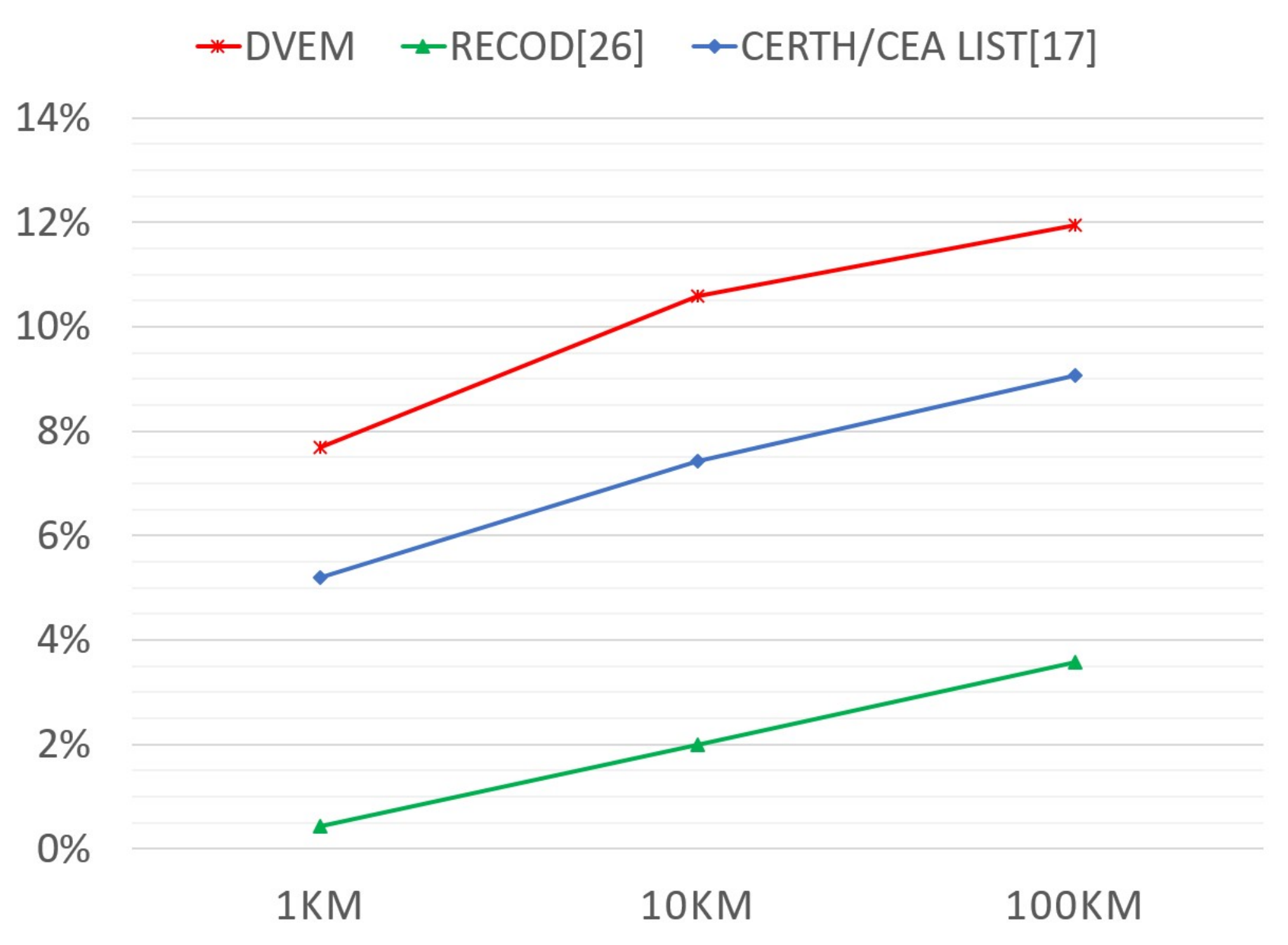}}
  \caption{HR@1 performance with respect to different evaluation radiuses on the MediaEval '15 Placing Task dataset.}\label{fig:res_ME15_HR1_stateOfArt}
\end{figure}

For geo-unconstrained location estimation, we compare our method to Li et al.~\cite{RECOD_MedEvl15}, and the neural network-based representation-learning approach by Kordopatis-Zilos~\cite{CERTH_MedEvl15}.
Results are reported on the entire test set as defined by the data release made by the MediaEval 2015 Placing Task.
The results in Fig.~\ref{fig:res_ME15_HR1_stateOfArt} show that our DVEM system redefines the state-of-the art on the MediaEval '15 dataset. 
Again, for completeness, we include additional discussion of our experimental design.
The submissions to the MediaEval 2015 Placing Task are not allowed to tune on the test data.
They do, however, have access to a leader board which includes $25\%$ of the test data.
In 2015, teams made a limited number of submissions to the leader board ($<=3$). 
Our experimental design was different in that we tuned on $2\%$ of the test data.
However, again referring back to Table~\ref{tab:Para} we can see the magnitude of the advantage that this choice gave us.
The worst parameter settings yielded performance that was lower than that of the best parameter settings by 1\% absolute.
If the performance of a very poorly tuned version of DVEM falls by this amount, it would still outperform its competitors in Fig.~\ref{fig:res_ME15_HR1_stateOfArt}.
We point out that the independence of the superiority of DVEM from the way in which the parameters are set can be considered a reflection of an observation already made above: the choice of the critical parameter $a$ is dependent on how data was captured in general (i.e., zoom-in vs zoom-out) and not on the specific composition of the dataset.

\section{Conclusion} \label{conclusion}
We have presented a visual-content-based approach for prediction of the geo-locations of images,
based on common sense observations about challenges presented by visual patterns in image collections
These observations led us to propose a highly transparent approach that represents locations using visual element clouds representing the match between a query and a location, and leveraging geo-distinctiveness.
The evaluation conducted on two publicly available datasets demonstrates that the proposed approach achieves performance superior to that of state-of-the-art approaches in both geo-constrained and geo-unconstrained location estimation. 

We close with two additional observations about the value of the DVEM approach moving forward.
A key challenge is that the distribution of image data used for geo-unconstrained location prediction is highly sparse over many regions.
This sparsity has led to the dominance of search-based approaches such as DVEM over classification approaches, already mentioned above.
An additional consequence, we expect, is that the search-based framework will remain dominant, and that new, deep-learning approaches will contribute features, as in~\cite{CERTH_MedEvl15}, which can enhance, but will not replace, DVEM.
Note that because DVEM calculates representations over a `contextual' image set, rather than the whole collection, it is not forced to pre-define locations of a particular scale. 
The result is that DVEM is able to apply geo-distinctiveness to predict the location of images on a continuous scale, limited only by the visual evidence present in the data set.

\bibliographystyle{abbrv}

\bibliography{xinchaoBib}

\end{document}